\documentclass[aps,pra,twocolumn,superscriptaddress,nofootinbibt,longbibliography]{revtex4-2}
\usepackage[utf8]{inputenc}
\usepackage{microtype}
\usepackage{mathtools}
\usepackage{booktabs}
\usepackage{amsfonts}
\usepackage{amsbsy}
\usepackage{physics}
\usepackage[dvipsnames]{xcolor}
\usepackage{caption}
\usepackage{soul}

\usepackage{tikz}
\usetikzlibrary{shapes.geometric}
\usepackage{subcaption}
\usepackage{algorithm}
\usepackage{siunitx}
\usepackage[noend]{algpseudocode}
\usepackage[normalem]{ulem}

\newcommand{\vvec}[1]{\boldsymbol{#1}}%

\definecolor{numexpcolor}{rgb}{0.6,0.6,0.6}

\usepackage{amsthm}

\newcommand{\arcc}[2]{\scalebox{0.9}{\{}#1, #2\scalebox{0.9}{\}}}

\begin{document}

\begin{abstract}
Distance measures provide the foundation for many popular algorithms in Machine Learning and Pattern Recognition. Different notions of distance can be used depending on the types of the data the algorithm is working on.
For graph-shaped data, an important notion is the Graph Edit Distance (GED) 
that measures the degree of (dis)similarity between two graphs in terms of the operations needed to make them identical. As the complexity of computing GED is the same as NP-hard problems, it is reasonable to consider \textit{approximate} solutions. 
In this paper we present a QUBO formulation of the GED  problem. This allows us to implement two different approaches, namely quantum annealing and variational quantum algorithms that run on the two types of quantum hardware currently available:  quantum annealer and gate-based quantum computer, respectively. 
Considering the current state of noisy intermediate-scale quantum computers, we base our study on proof-of-principle tests of their performance.

\end{abstract}

\title{Computing Graph Edit Distance with Algorithms on Quantum Devices}

\author{Massimiliano Incudini}
\affiliation{Dipartimento di Informatica, Universit\`a di Verona, Strada le Grazie 15 - 34137, Verona, Italy}

\author{Fabio Tarocco}
\affiliation{Dipartimento di Informatica, Universit\`a di Verona, Strada le Grazie 15 - 34137, Verona, Italy}

\author{Riccardo Mengoni}
\affiliation{CINECA, Casalecchio di Reno, Bologna, Italy}

\author{Alessandra Di Pierro}
\affiliation{Dipartimento di Informatica, Universit\`a di Verona, Strada le Grazie 15 - 34137, Verona, Italy}

\author{Antonio Mandarino}
\affiliation{International Centre for Theory of Quantum Technologies, University of Gda\'nsk, 80-308 Gda\'nsk, Poland}

\date{\today}
\maketitle

\section{Introduction}

The \emph{Graph Edit Distance} (GED) \cite{ged_first_paper, GEDSurvey} 
represents one of the most common dissimilarity measures used in Pattern Recognition and Image Processing \cite{apppatternmatching, conte2004thirty}.
It has been successfully applied to many real-world tasks such as image recognition, handwritten digit recognition, face and face expression recognition \cite{appocr,appfacerecognition,kisku2010face,mengoni2021facial,appfingerprint}, and has applications in a variety of areas from computer vision and bioinformatics \cite{appbioinfo} to cognitive science \cite{appneurology}, and hardware security \cite{apphardwaresec}.
It has also been used in Machine Learning in order to define more powerful kernels for support vector machines \cite{bellet2011}, and in combination with kernel machines for pattern recognition \cite{Book-GED-Kernel}.

In general, the notion of edit distance (originally introduced for strings and then extended to graphs and general structured data) is given in terms of the  operations that must be performed on a pair of data in order to make them identical, and represents a quantitative estimate of their dissimilarity. For example in pattern recognition it measures the strength of the distortions that have to be applied to transform a source pattern into a target pattern.
As computing the edit distance is essentially to search for the best (in terms of cost) set of operations in the space of all possible ones, the problem is intrinsically a combinatorial optimization problem and its complexity depends on the structure of the data in the search space.
When data are graphs, calculating the edit distance becomes easily inefficient as the complexity of the search space grows exponentially with the number of nodes of the graphs. In fact, the GED problem is an NP-hard optimization problem \cite{zeng2009comparing}, which makes exact approaches impossible to use for large graphs. This makes the study of approaches that give near-optimal results in reasonable time an urgent and essential one, also in the light of the practical impact of the GED problem. It is therefore of great importance to discuss and analyze in this context the feasibility of quantum approaches. 
Many classical algorithms have been proposed up to now \cite{GEDSurvey}, which however cannot be straightforwardly translated to the quantum setting.  

In this paper we introduce a formulation of the GED problem as a Quadratic Unconstrained Binary Optimization (QUBO) problem, and show that this offers a common base for a direct implementation of the problem on both quantum and classical hardware, 
thus making it easy to compare the performances of various approaches. 
Our formulation is based on a \emph{quantified} version of the graph isomorphism problem, where the expected answer is the actual number of operations that are needed to make two graphs isomorphic, rather than `yes' or `no'.

The quest for optimization algorithms that can run on noisy intermediate scale quantum (NISQ) computers is a very timely topic, as well as a challenging task \cite{Preskill2018, QCscience}. 
The two main quantum strategies that have been proposed to tackle optimization tasks are Quantum Annealing (QA) \cite{quantumannealing1998kadowaki} and the circuit-based variational approaches. 
In this paper we will assess the suitability of quantum annealing and two variational algorithms, i.e. Variational Quantum Eigensolver (VQE) \cite{vqeperuzzo} and Quantum Approximate Optimization Algorithms (QAOA) \cite{fahriqaoa}, on finding an approximate solution of the GED problem.

Moreover, we will give a concise comparative study of their performance with respect to a specific classical algorithm, namely Simulated Annealing, which makes use of the same QUBO formulation of the problem as the quantum strategies.  



\subsection{The Quantum Strategies}
The QA approach can be seen as the quantum analogue of the classical simulated annealing optimzation algorithm \cite{simulatedannealing1983}, and essentially consists in using a specific search procedure for  
finding the state of minimum energy of spin systems studied in statistical mechanics, 
e.g. the Ising model in a random field \cite{quantumannealing1998kadowaki}. In the quantum case the fluctuations needed to scan the energy landscape are provided by a field that allows quantum tunneling, in contrast to the simulated annealing where the fluctuations are thermal \cite{ReviewQannealing}. 
While QA is based on the adiabatic model of quantum computation, VQE and QAOA are hybrid quantum-classical algorithms implemented on circuit-based quantum computers. This approach has been tested in several proof-of-principle experiments for real-world optimization (e.g. scheduling) problems \cite{qaoa2020vikstaal, qaoapaper2}. 

Both kinds of approaches share a common physical ground, since they are both implemented by means of a total Hamiltonian
that is composed by two non-commuting terms, $H_0$ and $H_1$, such that $[H_0, H_1] = H_0H_1 - H_1H_0 \neq 0 $. This Hamiltonian has the form
\begin{equation}
\label{eq:genH}
     H = (1 - \lambda (t) ) H_0 + \lambda (t) H_1,
 \end{equation}
where $\lambda(t)$ is a control function that is valued in the interval $[0,1]$ and allows us to switch between the two terms. 
In fact, assuming that the computation time is in the interval $[0, \tau]$, one can initialize the system into the ground state of $H_0,$ imposing $ \lambda(0) = 0,$ and tune the control parameter till reaching $\lambda(\tau)=1.$ 
If the variation of $\lambda(t)$ is such that the hypotheses of the adiabatic theorem \cite{Katoadiabatic} are satisfied, both algorithms can be viewed as prefatory approaches for full adiabatic quantum computation \cite{Revadiabatic}. Moreover, the gates composing the QAOA can be seen as a Suzuki-Trotter decomposition \cite{mbeng2019quantum} of the unitary evolution stemming from a Hamiltonian as defined in (\ref{eq:genH}). 

Another common aspect of QA and QAOA is that their performance significantly depends on heuristics for the choice of the annealing schedule and parameter initialization, respectively.

By defining the problem of calculating the GED as a Quadratic Unconstrained Binary Optimization (QUBO) problem, 
we are able to experiment  both on quantum annealers, such as the D-Wave Systems Inc. machines, and on quantum circuits via variational algorithms such as {VQE and} QAOA. 
We introduce a QUBO formulation of GED that is inspired by the one presented in \cite{quboformulationgi} for the Graph Isomorphism problem, i.e. the computational problem of determining whether two finite graphs are isomorphic. Our formulation exploits the fact that GED can be seen as a quantitative generalization of  graph isomorphism, which requires a `counting' phase while checking the nodes/edges in each of the graphs. This allows us to obtain a quantitative answer to the optimization problem rather than just a yes/no answer as in the case of graph isomorphism. This also explains why the GED problem is far more complex than graph isomorphism, for which it was recently shown that the problem is solvable in quasi-polynomial time \cite{babai15}.

Our strategy is to use the QUBO formulation across all the platforms: classical, quantum annealers and gate-based quantum computers. This formulation of the problem, which is common for all the platforms, allows their smooth comparison and the identification of the most suitable one. 
The NISQ devices that we use to run the QUBO formulation of GED are the D-Wave System quantum annealer and the IBM circuit-based quantum computer.
This will contribute to assess the actual power of NISQ devices for real-world problems. A benchmark of quantum optimization algorithms on satisfiability and max-cut problems has been introduced in \cite{willsch2020benchmarking}. 
In contrast to that work, we analyze the GED problem, which requires a more involved QUBO formulation and therefore a larger number of qubits.
Moreover, we consider a broader class of quantum algorithms and include also classical algorithms in our comparison. 

This paper is structured as follows. In Sec.~\ref{sec:1prereq} we introduce the notion of graph edit distance, and in Sec.~\ref{sec:qubo_formulation} its QUBO formulation.
Section~\ref{sec:classical-approaches} is devoted to the classical approaches developed to compute GED, while in Section~\ref{sec:methods} we discuss the methods we employ 
in our experiments on both quantum annealer and circuit-model quantum computer. In Section~\ref{sec:numerical_results} we present the results of our experiments and provide a comparison among the different strategies. Finally, in Section~\ref{sec:conclusions} we discuss our results in the light of the currently available technologies and we address some open questions.

\section{Graph Edit Distance}\label{sec:1prereq}
A graph \(G = (V, E)\) consists of a finite, non-empty set of vertices \(V\) of cardinality $|V|=n$, and a set of edges \(E = \{ \arcc{u}{v} \mid u, v \in V\}  \subseteq V \times V \) of cardinality $|E| \le n^2$. The graph is \emph{undirected} if each edge is described by an unordered pair of vertices, \emph{directed} otherwise. In an undirected graph, the number of edges is at most $n(n+1)/2$ or, by excluding \emph{self-loops} (edges starting and ending in the same vertex), $n(n-1)/2$. 

Two graphs $G_1=(V_1,E_1)$ and $G_2=(V_2,E_2)$ are \emph{isomorphic}, denoted by $G_1 \cong G_2$, if there exists a \emph{bijection} (1-to-1 mapping) $\pi$ between the vertex sets of $G_1$ and $G_2$ satisfying the property of \emph{edge-preserving} \cite{fortin1996graph}, i.e. $\arcc{u}{v} \in E_1 \iff \arcc{\pi(u)}{\pi(v)} \in E_2$. 
The \emph{graph isomorphism} problem is the problem of establishing the exact matching of two graphs \cite{fortin1996graph}.
The inexact matching is the more general case where there is a difference between two graphs, and a measure of this difference quantifies `how much' the two graphs are (dis)similar. An important graph similarity measure is the \emph{graph edit distance} (GED) \cite{ged_first_paper, GEDSurvey}, whose value is the total cost of `transforming' one graph into the other, thus making them isomorphic. 
Clearly, when two graphs are isomorphic, the GED between them is zero.

The GED problem is  a NP-hard optimization problem, i.e. intuitively, it is at least as hard as the hardest problems in NP. This means that it is computationally more expensive than graph isomorphism for which a recent result shows that it is possible to find a solution in quasi-polynomial time \cite{babai15}. 

A \emph{graph edit operation} is a mapping from the set of graphs to itself. 
The most common edit operations on a graph $G=(V,E)$ are listed in Table \ref{tab:operation}, where we also specify the part of the graph which they act on (Vertex or Edge set), and the condition of their applicability (Constraint).
Specifically, these operations can be performed provided that they do not insert a pre-existing vertex/edge, or delete a non-existing vertex/edge; it must also be guaranteed that before deleting a vertex, each edge starting or ending into that vertex must be deleted.

\begin{table}[H]
    \begin{tabular}{lccl}
        \toprule
        Operation & Vertex set & Edge set & Constraint \\\midrule
        Insert node $v$ & $V \cup \{ v \}$ & No changes & $v \not\in V$ \\
        Delete node $v$ & $V \setminus \{ v \}$ & No changes & $v \in V, \not\exists uv \in E$\\
        Insert edge $uv$ & No changes & $E \cup \{uv\}$ & $uv \not\in E$\\
        Delete edge $uv$ & No changes & $E \setminus \{uv\}$ & $uv \in E$ \\ \bottomrule
    \end{tabular}
    \caption{Main operations composing a graph edit path}
    \label{tab:operation}
\end{table}

\begin{figure}[htbp]
    \begin{subfigure}[b]{0.5\textwidth}
        \scalebox{0.8}{\begin{tikzpicture}
            \node[regular polygon,regular polygon sides=5,minimum size=3cm] (p) {$G_1$};
            \foreach\x in {1,...,5}{\node[] (p\x) at (p.corner \x){\x};}
            \draw (p1) -- (p2) -- (p3) -- (p4) -- (p5) -- (p1);
            \node[regular polygon,regular polygon sides=5,minimum size=3cm, xshift=5cm] (q) {$G_2$};
            \foreach\x in {1,2,3,5}{\node[] (q\x) at (q.corner \x){\x};}
            \draw (q3) -- (q5) -- (q1) -- (q2) -- (q3);
        \end{tikzpicture}}
       \caption{}
        \label{fig:onepath_step_1}
    \end{subfigure} \\
    \begin{subfigure}[b]{0.5\textwidth}
        \scalebox{0.8}{\begin{tikzpicture}
            \node[regular polygon,regular polygon sides=5,minimum size=3cm] (p) {$P(G_1)$};
            \foreach\x in {1,2,3,5}{\node[] (p\x) at (p.corner \x){\x};}
            \foreach\x in {4}{\node[red] (p\x) at (p.corner \x){\x};}
            \draw (p1) -- (p2) -- (p3);
            \draw[red, dashed] (p3) -- (p4); 
            \draw[red, dashed] (p4) -- (p5);
            \draw[green, dashed] (p3) -- (p5);
            \draw (p5) -- (p1);
            \node[regular polygon,regular polygon sides=5,minimum size=3cm, xshift=5cm] (q) {$G_2$};
            \foreach\x in {1,2,3,5}{\node[] (q\x) at (q.corner \x){\x};}
            \draw (q3) -- (q5) -- (q1) -- (q2) -- (q3);
        \end{tikzpicture}}
       \caption{}
        \label{fig:onepath_step_2}
    \end{subfigure}
    \caption{Solution of an instance of GED problem. (a) Initial graphs. (b) The edit path $P$ consisting of four edit operation (insertion of $\arcc{3}{5}$, deletion of $\arcc{3}{4}, \arcc{4}{5}$, deletion of vertex 4) is such that $P(G_1) \cong G_2$. Moreover $\ell(P)$ is the GED since no shorter path exists that makes the graphs isomorphic.}
    \label{fig:onepath}
    \end{figure}
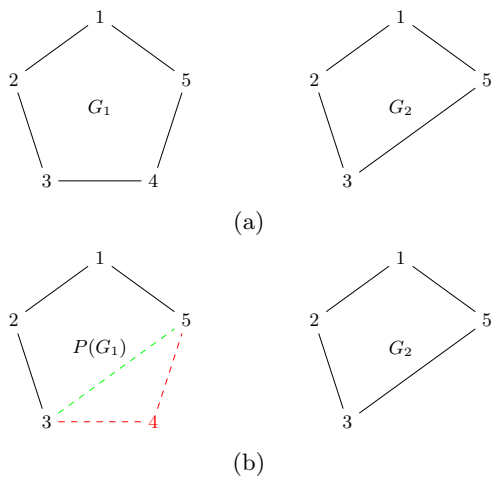

A \emph{graph edit path} $P$ is a composition of graph edit operations and the number of these operations defines the length, $\ell(P)$, of the path. The \emph{graph edit distance} between two graphs $G_1$ and $G_2$ can be defined as
\begin{equation}
    \text{GED}(G_1, G_2)= \min_P\{ \ell(P) \mid 
P(G_1) \cong G_2 \}.
\end{equation}
A more fine-grained definition of GED can be found in \cite{BlumenthalGED}, where \emph{labelled graphs} are considered, so that each graph edit operation might contribute to the graph edit distance with a different weight. For our purpose, it is sufficient to consider the  case of unlabelled graphs, which allows us to keep our implementation simpler.
In particular, in the calculation of the GED we will assume that each edit operation has cost $1$. This implies that the length of an edit path effectively corresponds to the number of operations composing it, which is a reasonable estimate of the complexity of the calculation.
An example of calculation of GED is shown in Figure~\ref{fig:onepath}.

\section{QUBO formulation of GED}\label{sec:qubo_formulation}
The graph matching problem can be encoded as a linear optimization problem \cite{linearapproxgraphmatching, linearbinaryged} or a quadratic optimization problem \cite{quadraticgraphmatching,quboformulationgi}. 
We show here how the second approach can be extended to inexact graph matching, and precisely to the GED problem. The idea is to reformulate GED as a Quadratic Unconstrained Binary Optimization (QUBO) problem, a class of problems that is well known in multivariate optimization. 
As the name suggests, a QUBO problem is defined in terms of a \emph{quadratic} function of \emph{binary} variables \(x_i\),
which is \emph{unconstrained} or, more precisely, with constraints replaced by penalty terms and encoded within a matrix \(Q\) representing the objective function \cite{qubosurvey}.
The task is to find the value $\mathbf{x}^*$ such that
\begin{equation}
  \mathbf{x}^* = \arg \min_{\mathbf{x}} \mathbf{x}^T Q \mathbf{x},
\end{equation}
where \(x_i \in \{0,1\}\) for any of the $n$ entries of $\mathbf{x}$, and $Q$ is an \(n \times n\) upper triangular real valued matrix with elements $q_{i,j}$ encoding the data specification. 

QUBO problems are NP-hard problems \cite{Pardalos92} and 
share a similar structure and the same computational complexity with spin glass Ising models. This similarity allows for an almost immediate implementation of QUBO objective functions into the adiabatic model of quantum computation.

A QUBO formulation of the GED problem can be given as follows. 
Given two graphs $G_1 = (V_1, E_1)$ and $G_2 = (V_2, E_2)$ we shall assume that the number of vertices is the same in both graphs, that is $|V_1| = |V_2| = n$. This is without loss of generality; in fact, if $G_1$ has $k = |V_1| - |V_2| > 0$ vertices more than $G_2$ we build $G_2'$ by adding $k$ isolated vertices to $G_2$, then we calculate he GED between $G_1$ and $G_2'$, and finally add $k$ to the given value for considering the $k$ insertion node edit operations. 
The formulation requires $n^2$ variables, each denoted by $x_{i, j}$, with $x_{i,j} = 1$ if and only if the $i$-th vertex of $G_1$ is mapped to the $j$-th vertex of $G_2$. 

\begin{figure}[htbp]
\begin{subfigure}[b]{0.5\textwidth}
    \scalebox{0.8}{\begin{tikzpicture}
        \node[regular polygon,regular polygon sides=5,minimum size=3cm] (p) {$G_1$};
        \foreach\x in {1,...,5}{\node[] (p\x) at (p.corner \x){\x};}
        \draw (p1) -- (p2) -- (p3) -- (p4) -- (p5) -- (p1);
        \node[regular polygon,regular polygon sides=5,minimum size=3cm, xshift=5cm] (q) {$G_2$};
        \foreach\x in {1,...,5}{\node[] (q\x) at (q.corner \x){\x};}
        \draw (q3) -- (q5) -- (q1) -- (q2) -- (q3) -- (q4);
    \end{tikzpicture}}
    \caption{}
     \label{fig:twopath_step_0}
\end{subfigure}

\begin{subfigure}[b]{0.5\textwidth}
    \scalebox{0.8}{\begin{tikzpicture}
        \node[regular polygon,regular polygon sides=5,minimum size=3cm] (p) {$P(G_1)$};
        \foreach\x in {1,...,5}{\node[] (p\x) at (p.corner \x){\x};}
        \draw (p1) -- (p2) -- (p3) -- (p4); 
        \draw[red, dashed] (p4) -- (p5);
        \draw[green, dashed] (p3) -- (p5);
        \draw (p5) -- (p1);
        \node[regular polygon,regular polygon sides=5,minimum size=3cm, xshift=5cm] (q) {$G_2$};
        \foreach\x in {1,...,5}{\node[] (q\x) at (q.corner \x){\x};}
        \draw (q3) -- (q5) -- (q1) -- (q2) -- (q3) -- (q4);
    \end{tikzpicture}}
    \caption{}
     \label{fig:twopath_step_1}
\end{subfigure}

\begin{subfigure}[b]{0.5\textwidth}
    \scalebox{0.8}{\begin{tikzpicture}
        \node[regular polygon,regular polygon sides=5,minimum size=3cm] (p) {$P_1(G_1)$};
        \foreach\x in {1,...,5}{\node[] (p\x) at (p.corner \x){\x};}
        \draw (p1) -- (p2) -- (p3) -- (p4) -- (p5) -- (p1);
        \draw[green, dashed] (p3) -- (p5);
        \node[regular polygon,regular polygon sides=5,minimum size=3cm, xshift=5cm] (q) {$P_2(G_2)$};
        \foreach\x in {1,...,5}{\node[] (q\x) at (q.corner \x){\x};}
        \draw (q3) -- (q5) -- (q1) -- (q2) -- (q3) -- (q4);
        \draw[green, dashed] (q4) -- (q5);
    \end{tikzpicture}}
    \caption{}
     \label{fig:twopath_step_2}
\end{subfigure}
    \caption{Equivalence between the standard and the insertion-only definition of GED between $G_1$ and $G_2$, with $GED=2$.  (a) The bijection used to calculate the $GED$, $\pi(i) = i \; \forall i \in  [1, 5]$. (b) A graph edit path $P, |P|=2$, such that $P(G_1) \cong G_2$ according to $\pi$. The insertion is denoted by a green dashed line and the deletion by a red dashed line. (c) Paths $P_1$ and  $P_2$ such that $|P_1|+|P_2|=2$ and $P_1(G_1) \cong P_2(G_2)$ according to $\pi$. The only insertion operations are denoted by a green dashed line.}
    \label{fig:twopath}
\end{figure}
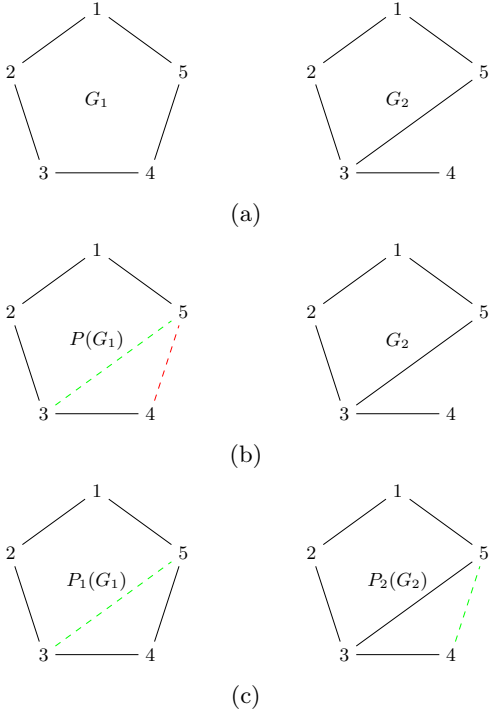

We define the dual path of $P$ as the path, $\tilde{P}$, where any deletion operation becomes an insertion operation and vice versa.
If a bijection $\pi$ corresponds to a graph edit path $P$ such that $P(G_1) \cong G_2$,  we can partition $P$ in two edit paths $P_1$ and $P_2$, where $P_1$ has only insertion operations and $P_2$ only deletion operations. Then
\[ P(G_1) \cong G_2 \text{ if and only if } P_1(G_1) \cong \tilde{P}_2(G_2). \]
Clearly, $\ell(P_1) + \ell(P_2) = \ell(P_1) + \ell(\tilde P_2) = \ell(P)$. Intuitively, the process is illustrated by the example reported in Figure~\ref{fig:twopath}. 

Considering the two assumption stated above, we can define the cost of a bijection $\pi$ as
\begin{equation}
\label{eqn:cost-bijection}
    \mathrm{cost}(\pi) = |E_1 \setminus \pi^{-1}(E_2)| + |E_2 \setminus \pi(E_1)|
\end{equation}
where $\pi(E) = \{\arcc{\pi(i)}{\pi(j)} \mid \arcc{i}{j} \in E \}$. For a given bijection $\pi$, the value of $\mathrm{cost}(\pi)$ is equal to the number of edges occurring in $E_1$ and not in $E_2$ 
plus the number of edges present in $E_2$ but not in $E_1$. 
The assumption that the graphs have the same number of vertices guarantees that a bijection exists. 

We can now define the GED as: 
\begin{equation}
\label{eqn:ged-restated}
\mathrm{GED}(G_1, G_2) = \min_{\pi} \mathrm{cost}(\pi).
\end{equation}

The QUBO formulation of our problem is composed by two parts:
a \emph{hard constraint} $Q_h$, whose function is to add a penalty if the solution $\vvec{x}$ does not represent a bijection (in that case, $\vvec{x}$ is not a solution for GED) and a \emph{soft constraint} $Q_s$, which introduces a penalty for any edge mismatch as in Equation~(\ref{eqn:cost-bijection}).

The hard constraint reads as 
\begin{align}
\label{eqn:hard-constraint}
Q_h(\vvec{x})  
& = \sum_{0 \le i < n}\underbrace{\Big(1 - \sum_{0 \le j < n} x_{i, j} \Big)^2}_{(a)} \nonumber \\
& + \sum_{0 \le j < n}\underbrace{\Big(1 - \sum_{0 \le i < n} x_{i, j} \Big)^2}_{(b)}
\end{align}
and ensures the validity of the solution (i.e. $\vvec{x}$ corresponds to a bijection). Whenever the $i$-th vertex of $G_1$ is mapped to zero or more than one vertices of $G_2,$ the $(a)$ term of Equation~(\ref{eqn:hard-constraint}) is greater than zero. The $(b)$ term of Equation~(\ref{eqn:hard-constraint}) is greater than zero, when the $j$-th vertex of $G_2$ is the image of none or more than one vertices of $G_1$.

The soft constraint is
\begin{align}
\label{eqn:soft-constraint}
Q_s(\vvec{x}) 
& = \sum_{\arcc{i}{j} \in E_1} R_{i,j}(\vvec{x}) \nonumber \\
& + \sum_{\arcc{i'}{j'} \in E_2} S_{i',j'}(\vvec{x})
\end{align}
with
\[ R_{i, j}(\vvec{x}) 
= \sum_{0 \le i' < n} x_{i, i'} \sum_{0 \le j' < n}  x_{j, j'} (1 - e_{i', j'}^{(2)})
\]
and 
\[ S_{i', j'}(\vvec{x}) 
= \sum_{0 \le i < n} x_{i, i'} \sum_{0 \le j < n}  x_{j, j'} (1 - e_{i, j}^{(1)})
\]
where 
\[e_{i,j}^{(k)} = \begin{cases} 1, \arcc{i}{j} \in E_k \\ 0, \arcc{i}{j} \not\in E_k \end{cases}.\]

The term $Q_s$ counts how many edges are not preserved by the bijection $\pi$ implied by $\vvec{x}$. In particular, the $R_{i,j}$ counts the arcs $\arcc{i}{j}$ in $G_1$, $\arcc{\pi(i)}{\pi(j)}$ missing in $G_2$. The $S_{i',j'}$ terms counts the arcs $\arcc{i'}{j'}$ in $G_2$, $\arcc{\pi^{-1}(i')}{\pi^{-1}(j')}$ missing in $G_1$. We prove now that the term $\sum_{i,j \in E_1} R_{i,j}(\vvec{x})$ is equivalent to the term $|E_1 \setminus \pi^{-1}(E_2)|$:
\begin{align*}
         & |E_1 \setminus \pi^{-1}(E_2)| \\
    = & \sum_{\arcc{i}{j} \in E_1} (1 - e_{\pi(i), \pi(j)}) \\
    = & \sum_{\arcc{i}{j} \in E_1} \left(x_{i, \pi(i)} x_{j, \pi(j)} (1 - e_{\pi(i),\pi(j)}) \right) \\
    = & \sum_{\arcc{i}{j} \in E_1} \left(\sum_{i' \le n} x_{i, i'} \sum_{j' \le n} x_{j, j'} (1 -  e_{i',j'}) \right) \\
    = & \sum_{\arcc{i}{j} \in E_1} R_{i,j}(\vvec{x})
\end{align*}
where $e_{ij}$ is one iff arc $\arcc{i}{j} \in E_2$. The procedure to identify $S_{i',j'}$ is equivalent. 
The complete formulation reads:
\begin{equation}\label{eqn:formulation}
    Q_{\lambda_h, \lambda_s}(\vvec{x}) = \lambda_h Q_h(\vvec{x}) + \lambda_s Q_s(\vvec{x})
\end{equation}
where the choice of parameter $\lambda_h, \lambda_s$ decides the weight of each penalty term. 

If we set $\lambda_h \gg \lambda_s$ we are guaranteed that all \emph{valid} solutions, i.e all those having null hard constraint contribution, have lower cost than any \emph{non-valid} solutions. For graphs of $n$ vertices, we can choose 
\begin{equation}
    \label{eqn:respecthardconstraints}
    \lambda_h > n^2 \lambda_s, 
\end{equation}
to ensure that the contribution of the soft constraint $\lambda_s Q_s(\vvec{x})$ to the total QUBO problem is always smaller than the contribution of hard one $\lambda_h Q_h(\vvec{x})$. This is because $n^2$ is the cost of the worst case, i.e when one graph is complete and the other one empty. 

If the maximum cardinality of edges is $|E|=m<n^2$, then 
Equation~(\ref{eqn:respecthardconstraints}) becomes $\lambda_h > m \lambda_s$. 

\section{Classical and quantum approaches to GED}\label{sec:classical-approaches}
Many classical algorithms for the GED problem exist \cite{BlumenthalGED, blumenthal2020}. Clearly, given the computational complexity of GED,
all known exact algorithms run in exponential time. A widely used  approach is based on the A* search algorithm \cite{gedexactastar}. 
A different approach is to consider classical heuristics that run in polynomial time in the size of the input, however they are not all suitable to treat QUBO problems. The two most common approaches are either based on reduction GED problem to LSAPE (Linear Sum Assignment Problem with Error-Correction) \cite{bougleux2017hungarian} or based on local search \cite{localsearch}.


A different approach is the Simulated Annealing (SA) heuristic \cite{simulatedannealing1983}. This is a probabilistic algorithm that minimizes multivariate binary objective functions and can be used to solve QUBO problems. Due to its physical background discussed later in Section~\ref{sec:methods}, it can be compared with the approaches presented in the following sections. SA can be seen as taking a random walk in the solution space according to a Markov chain parametrized by a temperature parameter $T$ \cite{mcgeoch2014adiabatic}. We choose SA as classical counterpart in our comparison because it can straighforwardly be used to solve QUBO problems. 


\subsection{Simulated Annealing}
SA works as shown in Algorithm~\ref{alg:SA} in Appendix \ref{appendix:algo}. When starting the algorithm the temperature $T$ is high, and solutions with higher energy are accepted with a probability that follows the Boltzmann distribution. The temperature decreases exponentially and SA has fewer chances to accept high energy solutions. When the temperature is zero, SA works in a gradient-descent fashion and will converge to a local minima. SA is restarted many times (shots), each from a different point of the state space chosen randomly. Finally, the energy of the solution is the single, lowest energy found in all the shots. 

\subsection{Quantum annealing}

Quantum Annealing (QA) 
\cite{quantumannealing1989apolloni, quantumannealing1998kadowaki} 
is a meta-heuristic search algorithm that can be used to tackle QUBO problems. This optimization technique finds its roots in a problem mutuated by statistical physics, namely the search of the minimum-energy state of a spin system exhibiting a glassy phase \cite{PARISI}. 

We briefly recall the notation and the physics underlying a class of spin systems, hereafter referred to as Ising-like model. The aim is to clarify the connection between the unconstrained quadratic problems and the search of the ground state of such a class of models.   

Let us introduce a classical spin variable $s_i$ that can take values $\pm 1$. The function describing the energy of a system of $N$ interacting spin disposed over a $d-$dimensional discrete lattice is the Hamiltonian: 
\begin{equation}
   \label{eq:Hisingcl}
   H(s) = - \sum\limits_{\langle i , j \rangle} J_{i,j} s_i s_j - \sum\limits_{i} h_i s_i
\end{equation}
where the $\langle i , j \rangle$ denotes that the sum is over all the first neighboring sites, the $J_{ij}$ are the couplings between two sites of the lattice and $h_i$ is the external magnetic field acting on each spin. 

A quantum version of the Hamiltonian in Equation~(\ref{eq:Hisingcl}) is obtained replacing suitably the binary variable $s_i$ with an \emph{ad hoc} Pauli matrix $\sigma_{i}^\alpha$  with $\alpha=\{x, y, z \}$:
\begin{equation}
\label{eq:Hisingq}
H(\sigma) = - \sum\limits_{\langle i,j \rangle} J_{i,j} \sigma^z_i \sigma^z_j - \sum\limits_{i=1}^N h_i \sigma^x.      
\end{equation} 

The problem of finding the ground state of a Hamiltonian describing an \emph{Ising Spin Glass} problem is NP-hard and how it relates to the solution of many NP-hard is reported in \cite{isingmanynpcomplete}.

The QUBO problem is closely related to the problem of finding the ground state of a Hamiltonian  written in terms of spin variable, upon introducing the transformation: 
\begin{equation}
\label{eq:qubotoising}
    x_i \leftrightarrow \frac{1 + \sigma^z_i}{2}
\end{equation}

Originally, QA was introduced in \cite{quantumannealing1989apolloni} as a quantum-inspired, classical algorithm. 
In contrast to the SA in which the fluctuations to explore the energy landscape are provided by the temperature parameter $T$, quantum annealing uses a transverse field coefficient $\lambda(t)$ called \emph{tunneling coefficient} that modulates the two terms of the Hamiltonian as in Equation~(\ref{eq:genH}).

The quantum annealing uses the term $H_0$ usually referred to as the \emph{transverse field Hamiltonian} that does not commute with the term $H_1$ in which the optimization problem is encoded. The non-commutativity of the two terms provides the fluctuations necessary to exploit the quantum tunneling, and allows us to escape from the local minima by tunneling through hills in the solution landscape \cite{mcgeoch2014adiabatic}. 
Tipically, when the quantum hardware is used, the system is in a superposition of all possible state, with probability amplitudes depending on $H(t)$. 

Quantum annealers usually have a certain number of qubits which are connected according to a given  topology. Logically, any QUBO variable is mapped to a qubit. Physically, it is possible that two variables linked by a quadratic term $J_{i,j}$ are not connected. This requires us to find a \emph{minor embedding} \cite{minorembeddingchoi2008}, thus a mapping of variables to physical qubits such that variables bound by quadratic terms are located to adjacent qubits. If this is not possible, additional qubits are required to represent a single variable. For this reason, to each variable is assigned one logical qubit but this can be mapped to more than one physical qubits. The task of finding a minor embedding  consists in searching a  minor of the graph associated with the hardware topology  which is  isomorphic to the one associated to the QUBO problem.  Minor embedding  is currently solved using heuristics, available on the D-Wave Ocean library. 

It may happen that the physical qubits associated with a single logical qubits are prepared in different states: some in $s=-1$ and others in $s=+1$. This phenomena is called \emph{chain break} and is resolved in post-processing using either \emph{majority vote technique} (the logical qubit has the most frequent value in the physical qubits assignment) or \emph{energy minimization} (the logical qubit has the value that minimize the energy). The latter requires multiple evaluation of the formulation).

\subsection{Variational quantum algorithms}

\begin{figure}[htbp]
    \centering
\includegraphics[]{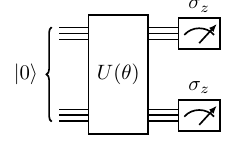}
    \caption{Parametric quantum circuit}
    \label{fig:vqe}
\end{figure}

We can use the QUBO formulation to solve the GED problem also on gate-based quantum computers, via Variational Hybrid algorithms, which are based on both classical and quantum resources. 

We define a parametric quantum circuit (PQC), that depends on an \emph{ansatz} for the values of the parameters $\vvec{\theta} = (\theta_1,..., \theta_M)$. The number of paramaters $M$ depends on the architecture of the circuit and the number of qubits $n$ (which are equivalent to the variables of the QUBO problem). For example, we can choose rotational gates which naturally depends on a set of rotation angles. As a shorthand we denote the composition of unitary gates composing the PQC by $U(\vvec{\theta})$, and the final state of such a circuit by $U(\vvec{\theta}) \ket{0}^{\otimes n} = \ket{\Psi(\vvec{\theta})}$. Then we compute the expectation value of the problem Hamiltonian $H_C$:
\begin{equation}
\label{eqn:expVar}
    E(\vvec{\theta}) 
    = \expval{H_C}{\Psi(\vvec{\theta})}.
\end{equation}
Such Hamiltonian is related to the QUBO formulation by Equation~(\ref{eq:qubotoising}). The value $E(\vvec{\theta})$ corresponds to the cost function which has to be minimized and the minimization stage is performed classically. 

Examples of this kind of algorithms are the variational quantum eigensolver (VQE) \cite{vqeperuzzo} and the quantum approximate optimization algorithm (QAOA) \cite{fahriqaoa}. They are mostly used to find the ground state of the Hamiltonian of nonintegrable spin systems \cite{theoryofvariationalhybrid}, which is indeed a minimization task. 

The Variational Quantum Eigensolver (VQE) is inspired by the Variational Principle \cite{vqeperuzzo} and it has found its most groundbreaking application in chemical-physics simulation \cite{VQE1,grossi1, grossi2},  and depending on the nature of the problem under investigation many different \emph{ans\"atze} can be used \cite{QChem_rev}. 
The circuit implementing it is shown in Figure~\ref{fig:varansalz}, and its construction is detailed in Algorithm~\ref{alg:vqeansatz} in Appendix~\ref{appendix:algo}.

\begin{figure}[htbp]
    \centering
    \scalebox{0.8}{\includegraphics[]{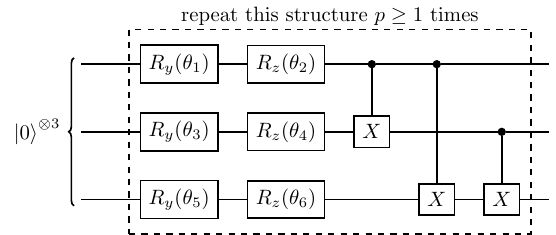}}
    \caption{Circuit for the variational ansatz used for VQE algorithm defined on $n=3$ qubits. The structure can be repeated $p$ times and the number of parameters $\theta$ is $2pn$.}
    \label{fig:varansalz}
\end{figure}

The Quantum Approximate Optimization Algorithm (QAOA) recently introduced in \cite{fahriqaoa, qaoa2020vikstaal, qaoapaper2, zhou2020quantum} is an application of VQE. The ansatz must respect a particular structure which depends on the Hamiltonian defining the problem. 

\section{Experimental setup}\label{sec:methods}

We consider a set of graphs with a number of vertices ranging from three to nine, and we randomly generate the graphs following the procedure explained in \cite{fastrandomgraphgeneration}. The procedure needs as inputs the number of vertices $n$ and the probabilities $p$ of generating edges in the graph.
For all the configurations labelled by the number of vertices $n \in  \{ 3, 4, ..., 9 \}$, we generate four graphs $G_1, G_2, G_3, G_4$ with edge probabilities $p_1 = 0.1, p_2 = 0.33, p_3 = 0.66, p_4 = 0.99$. 
This means that graph $G_1,$ generated with probability $p_1$, will contain a small number of edges while the graph $G_4$ is almost always a complete graph. 


We proceed now to exactly compute the $\mathrm{GED}(G_i, G_j)$ with $i, j= {1,2,3,4}$ for each number $n$ of vertices by means of the A* algorithm \cite{gedexactastar}. Note that this is feasible only for small sized graphs due to the exponential worst case complexity of the algorithm.
Any computational run has four inputs: the first graph $G_i$, the second graph $G_j$ and the parameters $\lambda_h, \lambda_s$ of Equation~\ref{eqn:formulation}. 
We iterate the runs over all possible pairs of graphs with the same number of vertices, including the pair of a graph with itself. 

The formulation shown in Equation~(\ref{eqn:formulation}) requires to set some values for parameters \(\lambda_h, \lambda_s > 0\). 
Experimental results, obtained by running SA on all pairs of values $\lambda_h=1$ and $\lambda_s \in \{1/i \mid i = 1, ..., 10\} \cup \{ 0.05, 0.01 \}$ on graphs up to $9$ vertices, show that the actual best value for $\lambda_s$ fulfilling Equation~(\ref{eqn:formulation}) would be $\lambda_s < \lambda_h/81 \approx 0.012$. 
Our choice of setting $\lambda_h=1$  and $ \lambda_s=0.1$ is a compromise between the need to fulfill Condition~(\ref{eqn:respecthardconstraints}) - which is valid for small $\lambda_s$ - and the the need to avoid values, $\lambda_s \simeq 0,$ which are so low as to make the soft constraint irrelevant.

We have run SA 1000 times, and kept the lowest energy  solution $\vvec{\tilde{x}}.$
For each run we have calculated the exact GED $s$ and the approximated one $\tilde{s}$. As the \emph{absolute error} $|s - \tilde{s}|$, grows with the problem size, we have introduced another quantity called the \emph{relative difference}, which is defined as
\begin{equation}
\label{eq:relativedifference}
    \Delta = \begin{cases}
    0, & s = \tilde{s} \\
    \left|\frac{\tilde{s} - s}{\max\{s, \tilde{s}\}}\right|, & s \neq \tilde{s}
\end{cases}
\end{equation}
and lies in the range $[0,1]$. 
The results of our experiments are shown in Table~\ref{tab:abchoice}. They suggest that the best values are those with $\lambda_s \in [1/4, 1/10]$. 
Choosing values of $\lambda_h$ and $\lambda_s$, such that the condition~(\ref{eqn:respecthardconstraints}) is satisfied, ensures the existence of a valid solution, however this is only a sufficient condition. In fact, we have strong numerical evidences that valid solutions exist also for other choices of parameters, for which a constraint linking $\lambda_h$ and $\lambda_s$ is still an open question. It is arguable that the best choice of parameters for SA will not necessary be the best one for quantum annealers and QAOA. However, notice that this choice is guided by the considerations explained above. The results of the experiments shown in Table \ref{tab:abchoice} empirically show the effectiveness of our choice.

\begin{table}[H]
    \centering
    \scalebox{0.9}{\begin{tabular}{crrrrrrrrrrrr}
    \toprule & 
    \multicolumn{12}{c}{Values of $\lambda_s$} \\
    $n$ & 1/1 & 1/2 & 1/3 & 1/4 & 1/5 & 1/6 & 1/7 & 1/8 & 1/9 & 0.1 & 0.05 & 0.01
    \\\midrule
    3  & 0.00 & 0.00 & 0.00 & 0.00 & 0.00 & 0.00 & 0.00 & 0.00 & 0.00 & 0.00 & 0.00 & 0.00 \\
    4  & 0.42 & 0.00 & 0.00 & 0.00 & 0.00 & 0.00 & 0.00 & 0.00 & 0.00 & 0.00 & 0.00 & 0.00 \\
    5  & 0.52 & 0.06 & 0.06 & 0.06 & 0.06 & 0.06 & 0.06 & 0.06 & 0.06 & 0.06 & 0.06 & 0.06 \\
    6  & 0.53 & 0.10 & 0.00 & 0.00 & 0.00 & 0.00 & 0.00 & 0.00 & 0.00 & 0.00 & 0.00 & 0.00 \\
    7  & 0.54 & 0.32 & 0.05 & 0.05 & 0.05 & 0.05 & 0.05 & 0.05 & 0.05 & 0.05 & 0.05 & 0.09 \\
    8  & 0.48 & 0.48 & 0.17 & 0.00 & 0.00 & 0.00 & 0.00 & 0.00 & 0.00 & 0.00 & 0.09 & 0.15 \\
    9  & 0.62 & 0.50 & 0.38 & 0.00 & 0.00 & 0.00 & 0.00 & 0.00 & 0.01 & 0.00 & 0.09 & 0.24 \\\bottomrule
    \end{tabular}}
    \caption{Average relative difference for the experiments with $\lambda_h=1, \lambda_s \in \{1/i \mid i = 1, ..., 10\} \cup \{ 0.05, 0.01 \}$ with respect to the number of vertices $n$.}
    \label{tab:abchoice}
\end{table}

The experimental setup is based on state-of-the art, freely available software. This approach facilitates reproducibility and avoids reinventing the wheel issues e.g. bugs, non-standard implementations.
As argued in \cite{mauerer2022123}, reproducibility is a serious issue in quantum science. We have addressed the problem by releasing the code in \url{github.com/incud/GraphEditDistance}. However, quantum annealer based experiments require the access to D-Wave machines which is available through the cloud.

\subsection{Classical approaches}

The exact calculation of GED is performed through the A* algorithm \cite{gedexactastar} implemented in NetworkX \cite{networkx}, 
while as heuristic  we use the 
SA implementation provided by D-Wave Ocean SDK. 

The control parameter in this implementation are set as follows. The temperature is initialized at $T_\text{start} = M/\log 2$ and decreases exponentially to $T_\text{end} = m/\log 100$, where $M$ and $m$ are an upper bound and a lower bound of the solutions, respectively. The (absolute value of the) smallest entry of the QUBO matrix is the lower bound $m$. The sum of the (absolute value of the) entries of the QUBO matrix is the upper bound $M$. 

Another implementation of SA is the one provided within the GEDLib library \cite{blumenthalgedlibpp} and its Python wrapper GEDLibPy \footnote{\texttt{https://github.com/Ryurin/gedlibpy}}. The comparison between this and the one provided by D-Wave is reported in Appendix~\ref{appendix:results} and shows that the results are comparable. 
We have also tested other state-of-the-art heuristics that are not based on QUBO formulation, however 
in general SA is the best performing (see Appendix~\ref{appendix:results}).
However we will rule out these heuristics from our comparison with the quantum algorithms and restrict ourselves only to methods specifically designed for QUBO problems. 


\subsection{D-Wave quantum annealers}

For our experiments we have used three different, (fully) quantum annealers from hardware company D-Wave: (i) D-Wave 2000Q; (ii) D-Wave Advantage 1.1; (iii) D-Wave Leap Hybrid Solver. We recall briefly their different characteristics. 

D-Wave 2000Q has 2041 qubits and uses Chimera topology having 5974 couplers (physical connections between qubits) \cite{dwave2000Q6}. 
D-Wave Advantage 1.1  has 5436 qubits and uses a Pegasus topology having 37440 couplers \cite{dwaveadv11}. Chain breaks are resolved using \emph{majority vote}, which is the cheapest technique and is the one suggested by the hardware vendor. D-Wave Leap Hybrid Solver is not a quantum annealer but a hybrid classical-quantum software whose configurations are fully managed by D-Wave and it is not customizable. 

Both D-Wave 2000Q and Advantage allow us to set the configuration of the annealing process, controlled by a parameter $s$ that grows monotonically from $0$ to $1$. The parameter is defined as $s=t/\tau$, where $\tau$ is the annealing time. For a specified \emph{annealing time} the evolution proceeds linearly in $t$. 

Due to the enhanced topology, it is expected to find more compact embedding with the latter hardware \cite{dwaveadvantagetechreport}.


Any GED computation runs on both machines with the following configurations:
\begin{enumerate}
    \item \emph{default configuration (DC)}: the annealing time is $20\si{\micro\second}$, the annealing process proceeds linearly;
    \item \emph{configuration Short Time (ST)}: the annealing time is $1\si{\micro\second}$, the annealing process proceeds linearly;
    \item \emph{configuration Long Time (LT)}: the annealing time $500\si{\micro\second}$, the annealing process proceeds linearly;
    \item \emph{configuration Pause Middle (PM)}: annealing schedule is 
    \begin{enumerate}
        \item in the first $5\si{\micro\second}$, $s$ grows from $0$ to $0.45$,
        \item in the following $94\si{\micro\second}$, $s$ remains constant,
        \item in the last $1\si{\micro\second}$, $s$ grows from $0.45$ to $1$.
    \end{enumerate}
    According to \cite{marhall2019powerofpausing}, the performance of pausing within the annealing schedule depends uniquely on the length of the pause (longer is better) and are independent of the annealing process before and after the pause.
\end{enumerate}

A run is the number of times the annealing is restarted. We have tried each configuration with a different number of runs: ST with $10^4$, DC with both $10^3$ and $10^4$, the others with $10^3$ and we have kept only the lowest energy result. 

We expect that the long time configuration might outperform those having shorter annealing time, due to the possibility of exploring the energy landscape widely. However, for longer times decoherence effects due to noise may arise and there is still no direct control of it. This increases the probability of errors.

We also expect the PM configuration to outperform the default configuration, due to the previous evidences suggesting that pausing the schedule in the middle of the process improves the performances \cite{marhall2019powerofpausing, improvingqapausing1, improvingqapausing2}.

\subsection{Variational Quantum Algorithms}

We used the implementation of VQE and QAOA available on the Qiskit IBM platform \cite{Qiskit}.

\begin{figure}[htbp]
    \centering
    \scalebox{0.75}{
\includegraphics[]{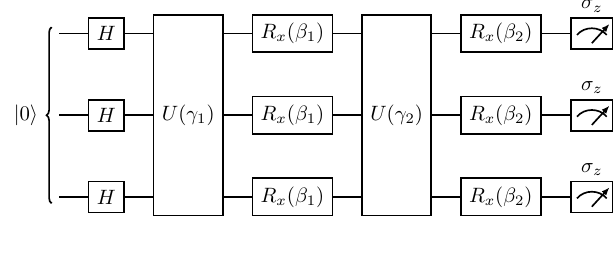}}
    \caption{Scheme of the QAOA circuit for $n=3$ qubits. The variational ansatz is repeated twice. The number of parameters is fixed to $2p$ (and does \emph{not} depends on $n$).}
    \label{fig:qaoacirc}
\end{figure}

We briefly recall the construction of the circuit implementing the QAOA as in \cite{fahriqaoa} and in Figure~\ref{fig:qaoacirc} we plot the circuit used to tackle the GED problem. The circuit has as input the state $\ket{+}^{\otimes n}$ where $n$ is the number of variables of the QUBO problem. Such state is obtained from $\ket{0}^{\otimes n}$ by applying the Hadamard gate to each qubit. The QAOA ansatz is constructed by repeating $p$ times two unitary operation $U_C(\gamma), U_B(\lambda_s)$. The whole ansatz will depend on $2p$ parameters $\gamma_1, \beta_1, ..., \gamma_{2p}, \beta_{2p} \in [-\pi, \pi]$. The first unitary operator is defined to be $U_C(\gamma) = \exp\{-i\gamma H_C\}$ where $H_C$ is the problem Hamiltonian and depends on our input. The mapping between the QUBO formulation and the Hamiltonian formulation is detailed in Equation~(\ref{eq:qubotoising}). The second unitary operator is called \emph{mixing Hamiltonian} and is defined to be $U_B(\beta) = \exp\{-i\beta H_B\}$, where $H_B = \sum_i \sigma^x_i$. Note that the two Hamiltonians $H_C, H_B$ must not commute to have not trivial results. The circuit gives as output the state
\begin{equation}
\label{eqqaoa}
\ket{\vvec{\gamma}, \vvec{\beta}} = U_B(\beta_p) U_C(\gamma_p) \cdots U_B(\beta_1) U_C(\gamma_1) \ket{+}^{\otimes n}. 
\end{equation}

The expectation value represents the energy (cost) associated to a particular choice of parameters and must be minimized:
\begin{equation} \label{energy_qaoa}
    E(\vvec{\gamma}, \vvec{\beta}) 
    = \expval{H_C}{\vvec{\gamma}, \vvec{\beta}}, 
\end{equation}
and we stress that the above equation is Equation~(\ref{eqn:expVar}) rewritten for the QAOA case.
Finally, the task of the classical optimazion is to find the optimal variational parameters such that: 
\begin{equation}
    (\vvec{\gamma}^*, \vvec{\beta}^*) = \arg \min_{\vvec{\gamma}, \vvec{\beta} } E(\vvec{\gamma}. \vvec{\beta}) 
\end{equation}
We perform 2048 runs with randomly initialized parameters, then the classical optimization is performed using the algorithm COBYLA \cite{cobyla}.
The construction of QAOA ansatz is detailed in Algorithm~\ref{alg:qaoaansatz} in Appendix~\ref{appendix:algo}.

\section{Numerical Results}\label{sec:numerical_results}

\begin{figure*}[t]
    \centering
    \begin{subfigure}[h]{0.30\textwidth} 
        \centering
        \includegraphics[width=\textwidth]{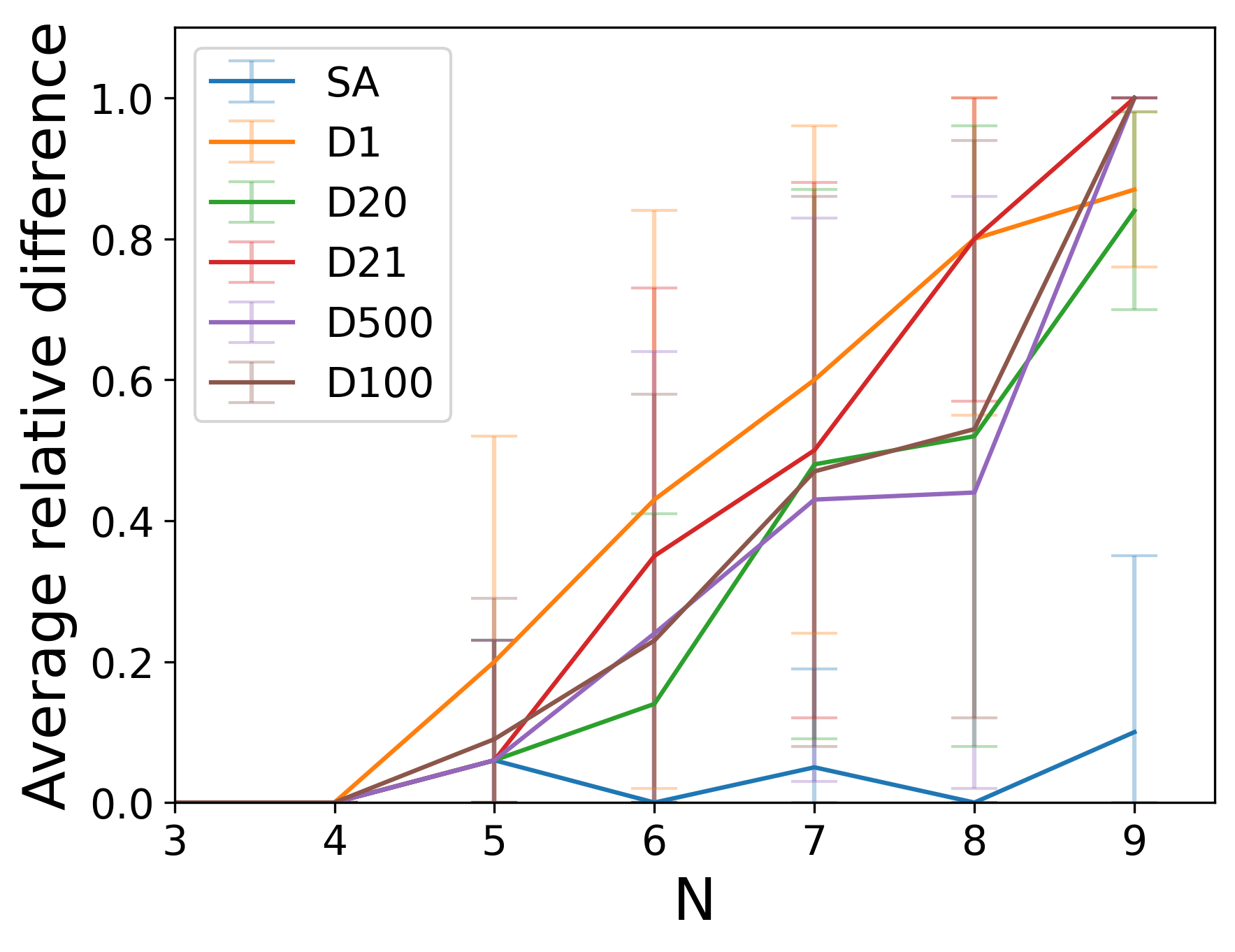}
        \caption{}
        \label{fig:result_dwave_2000}
    \end{subfigure}~\begin{subfigure}[h]{0.30\textwidth} 
        \centering
        \includegraphics[width=\textwidth]{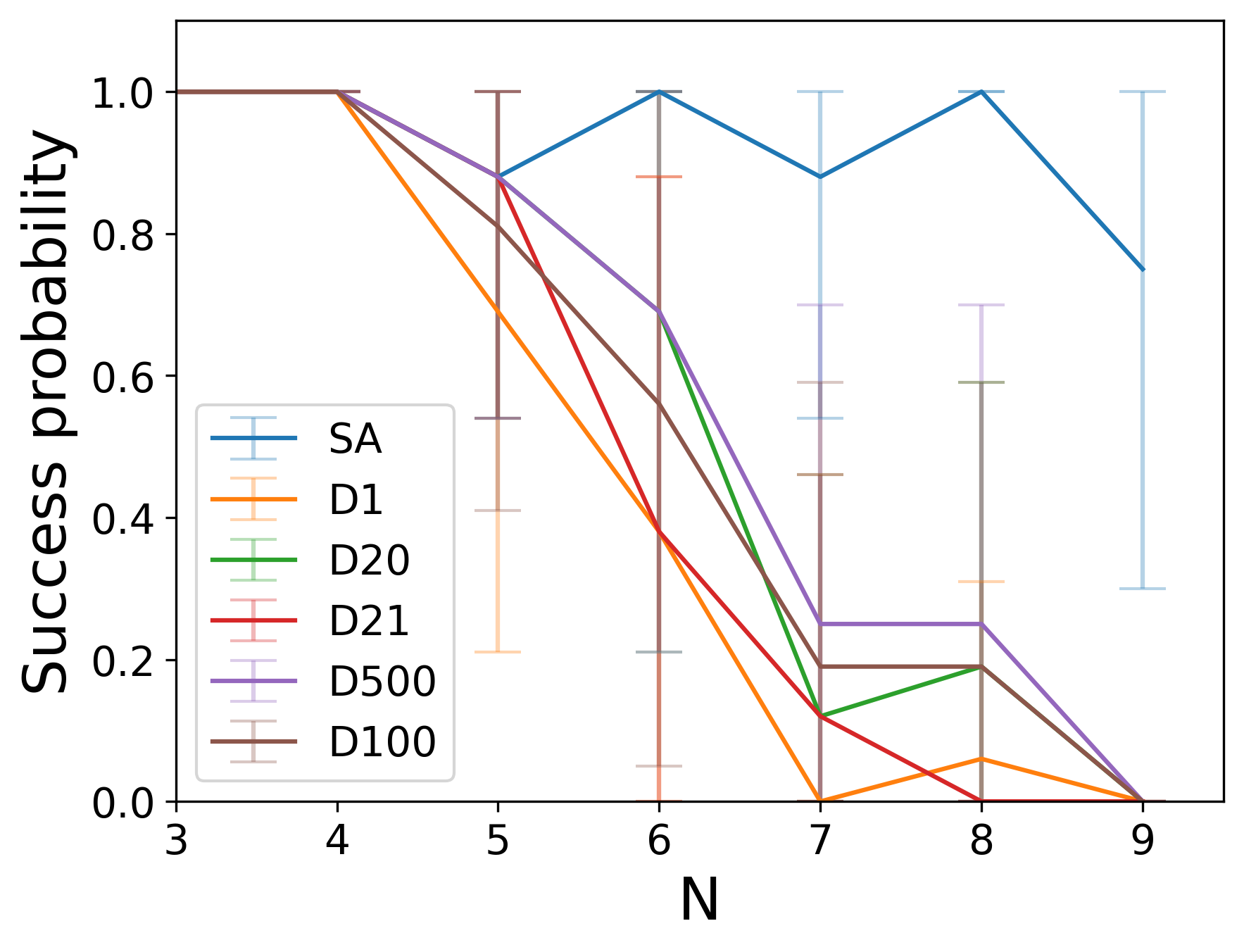}
        \caption{}
        \label{fig:success_prob_result_dwave_2000}
    \end{subfigure}~\begin{subfigure}[h]{0.30\textwidth} 
        \centering
        \includegraphics[width=\textwidth]{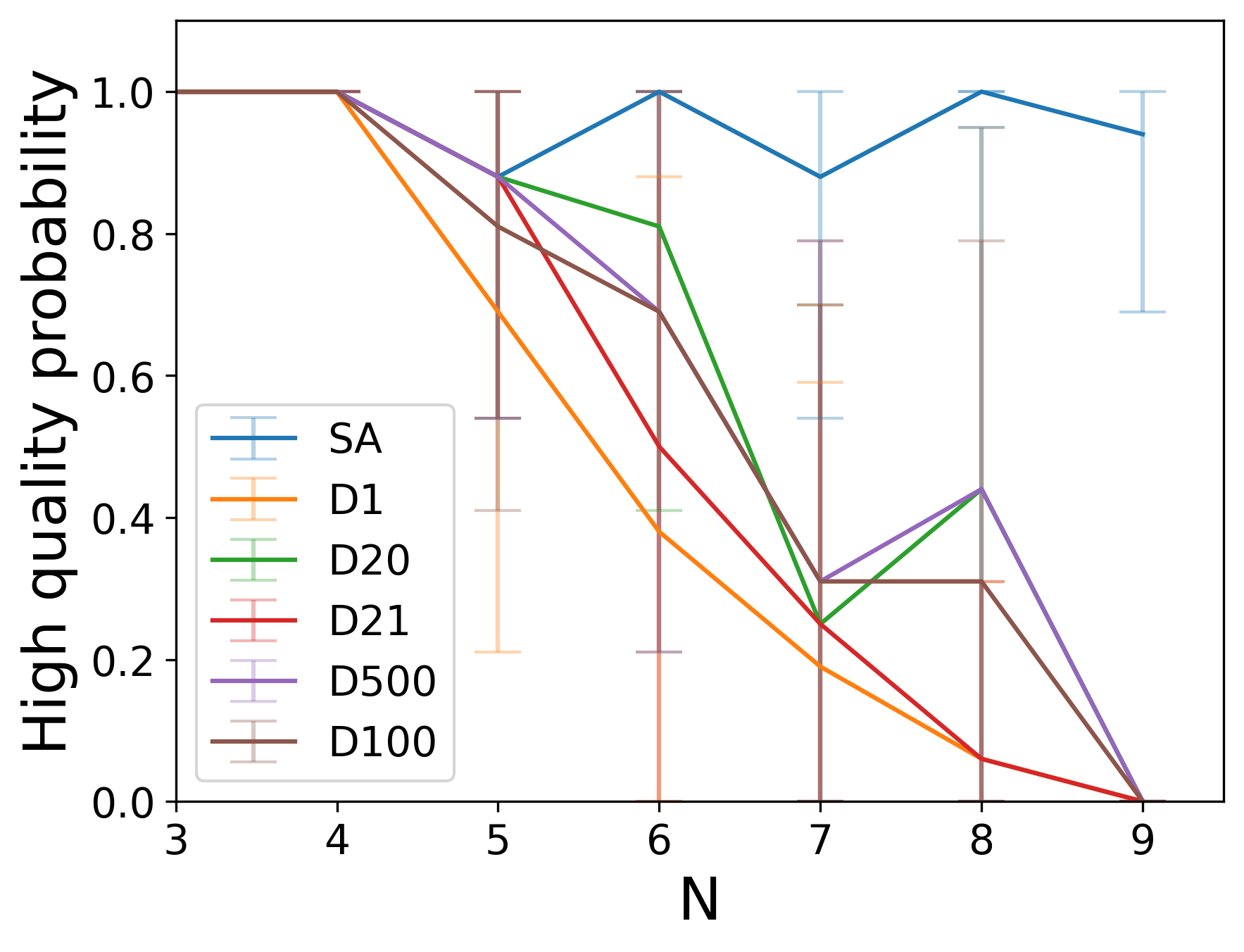}
        \caption{}
        \label{fig:hq_prob_result_dwave_2000}
    \end{subfigure} \\
    \begin{subfigure}[h]{0.30\textwidth} 
        \centering
        \includegraphics[width=\textwidth]{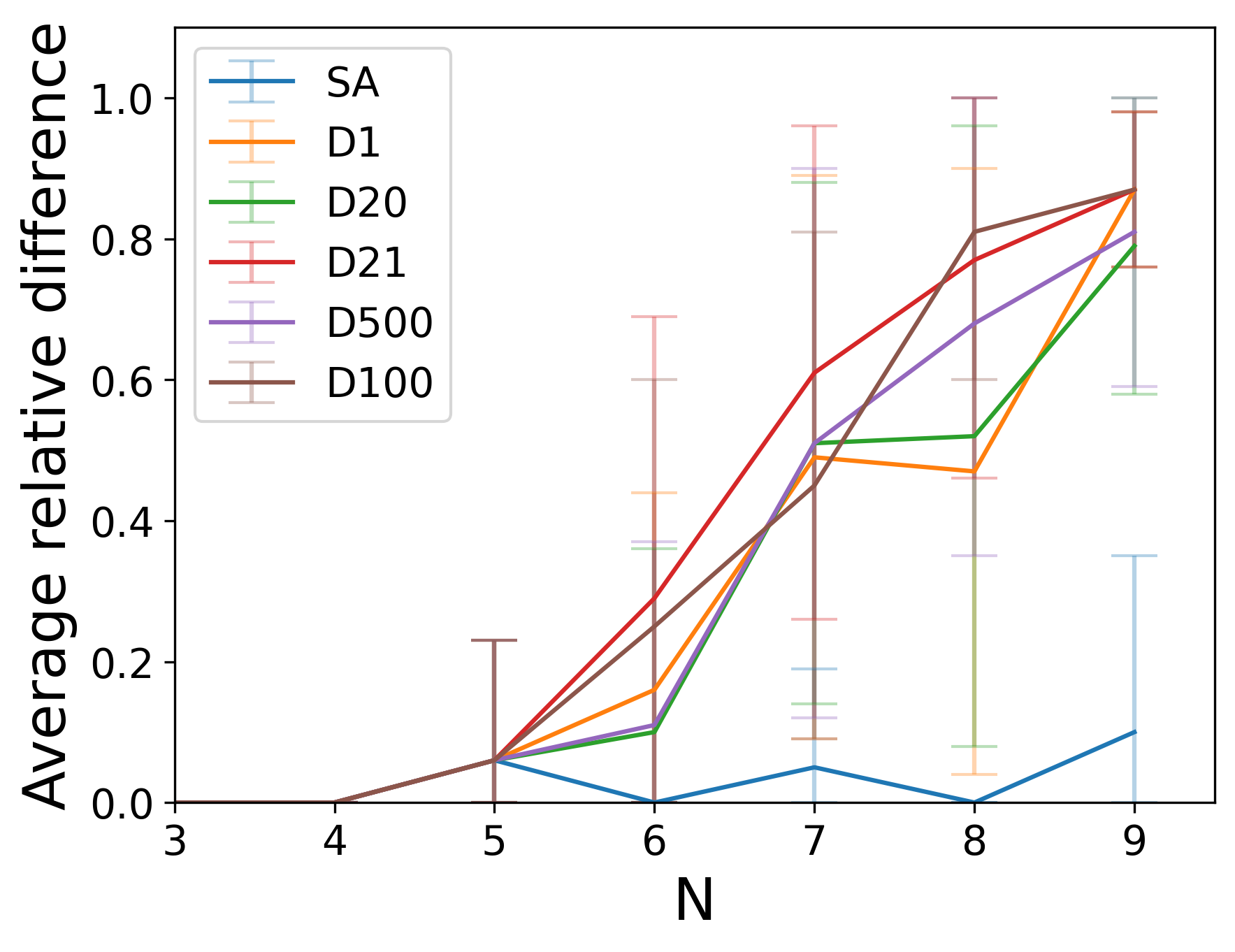}
        \caption{}
        \label{fig:result_dwave_advantage} 
    \end{subfigure}~\begin{subfigure}[h]{0.30\textwidth} 
        \centering
        \includegraphics[width=\textwidth]{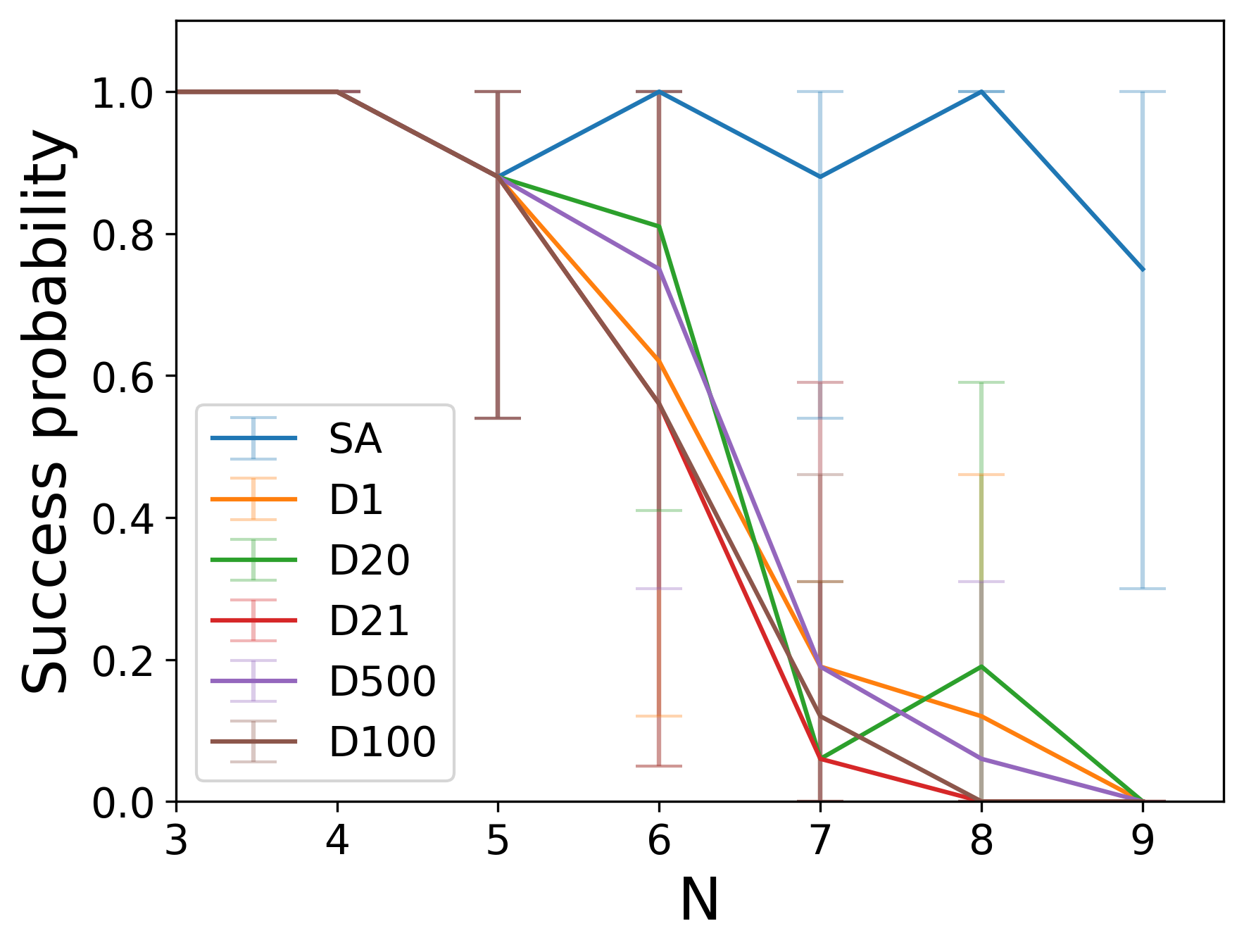}
        \caption{}
        \label{fig:success_prob_result_dwave_advantage}
    \end{subfigure}~\begin{subfigure}[h]{0.30\textwidth} 
        \centering
        \includegraphics[width=\textwidth]{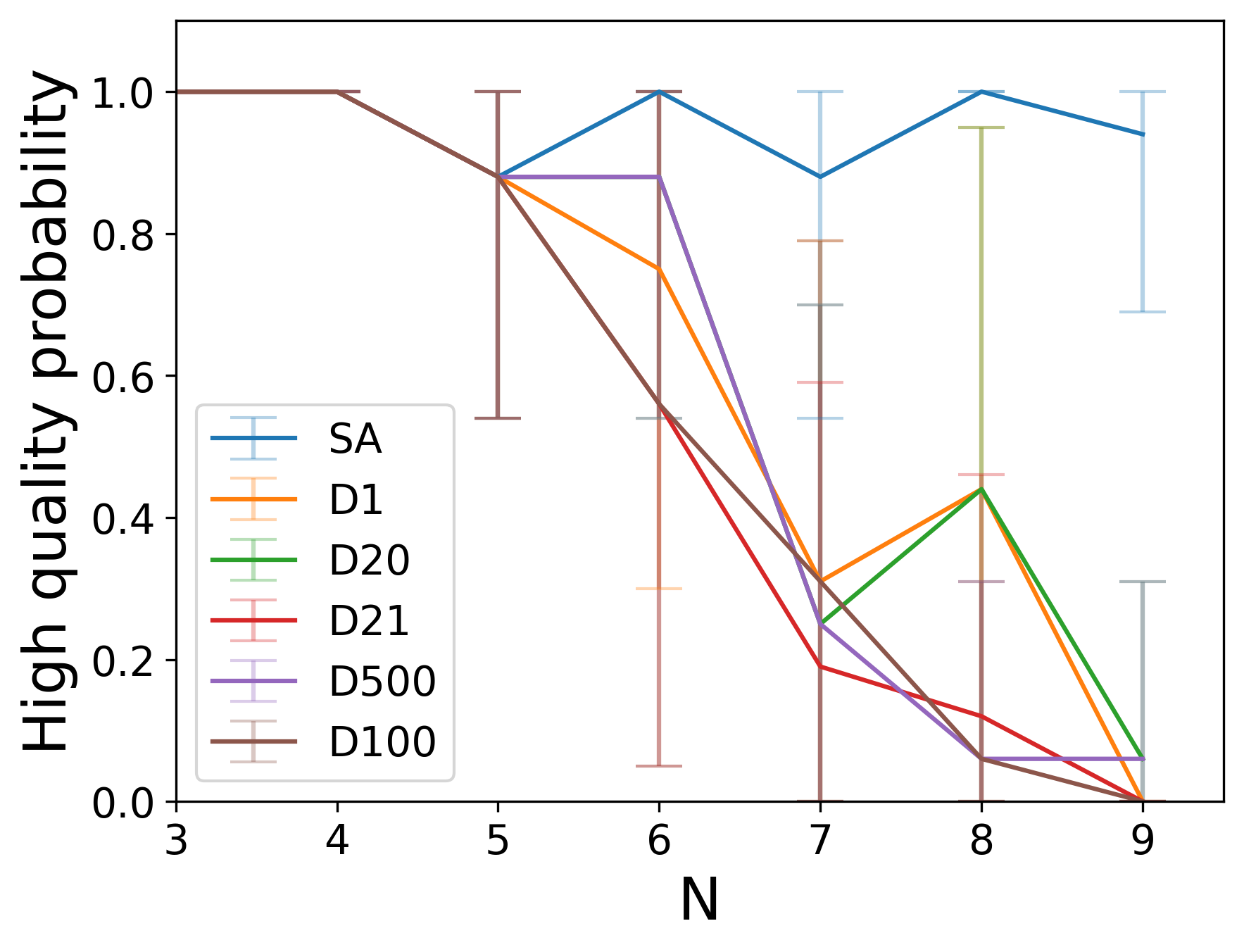}
        \caption{}
        \label{fig:hq_prob_result_dwave_advantage}
    \end{subfigure} 
    \caption{Comparison of the performance of the many configurations of quantum annealers as function of $n$ the number of vertices of a pair of graphs. Plots (a)-(b)-(c) compare configurations of D-Wave 2000Q, while (d)-(e)-(f) compares Advantage hardware. Plots (a)-(d) show the relative difference, (b)-(e) the success probability, (c)-(f) the high quality probability. The error bars represent the standard deviation. Legend:
    SA is simulateed annealing,
    D1 is annealing $1\si{\micro\second} \times 10^4$ runs, 
    D20 is annealing $20\si{\micro\second} \times 10^4$ runs, 
    D21 is annealing $20\si{\micro\second} \times 10^3$ runs, 
    D500 is annealing $500\si{\micro\second} \times 10^3$ runs, 
    D100 is annealing $100\si{\micro\second}$ paused in the middle and $\times 10^3$ runs}
    \label{fig:best_configuration}
\end{figure*}

Our analysis takes into account the number of vertices $n$, the hardware and its configurations. We evaluate the performance of each algorithm, illustrated in Appendix~\ref{appendix:algo}, using three different measures: the average of the \emph{relative difference} $\Delta$, 
the average of \emph{success probability}, and the average of \emph{high-quality probability}. The relative difference was defined in Equation~(\ref{eq:relativedifference}); the success probability is defined as the percentage of experimental results having $\Delta = 0$, and the
\emph{high-quality probability} is defined as the percentage of results having $\Delta \le 0.2$. 

All experiments uses the QUBO formulation shown in Equation~(\ref{eqn:formulation}) with parameters $\lambda_h=1, \lambda_s=0.1$, which are optimal according to the preliminary analysis in Table~\ref{tab:abchoice}. The choice of these values enforces the hard constraint, maximizing the chances of reaching a valid solution.

We summarized our results in Figures~\ref{fig:best_configuration}-\ref{fig:best_annealer}-\ref{fig:results_variational}. All the numerical results are reported in Appendix~\ref{appendix:results}.

\subsection{Resource usage}

\begin{figure*}[t]
    \centering
    \begin{subfigure}[h]{0.26\textwidth}
        \centering
        \includegraphics[width=\textwidth]{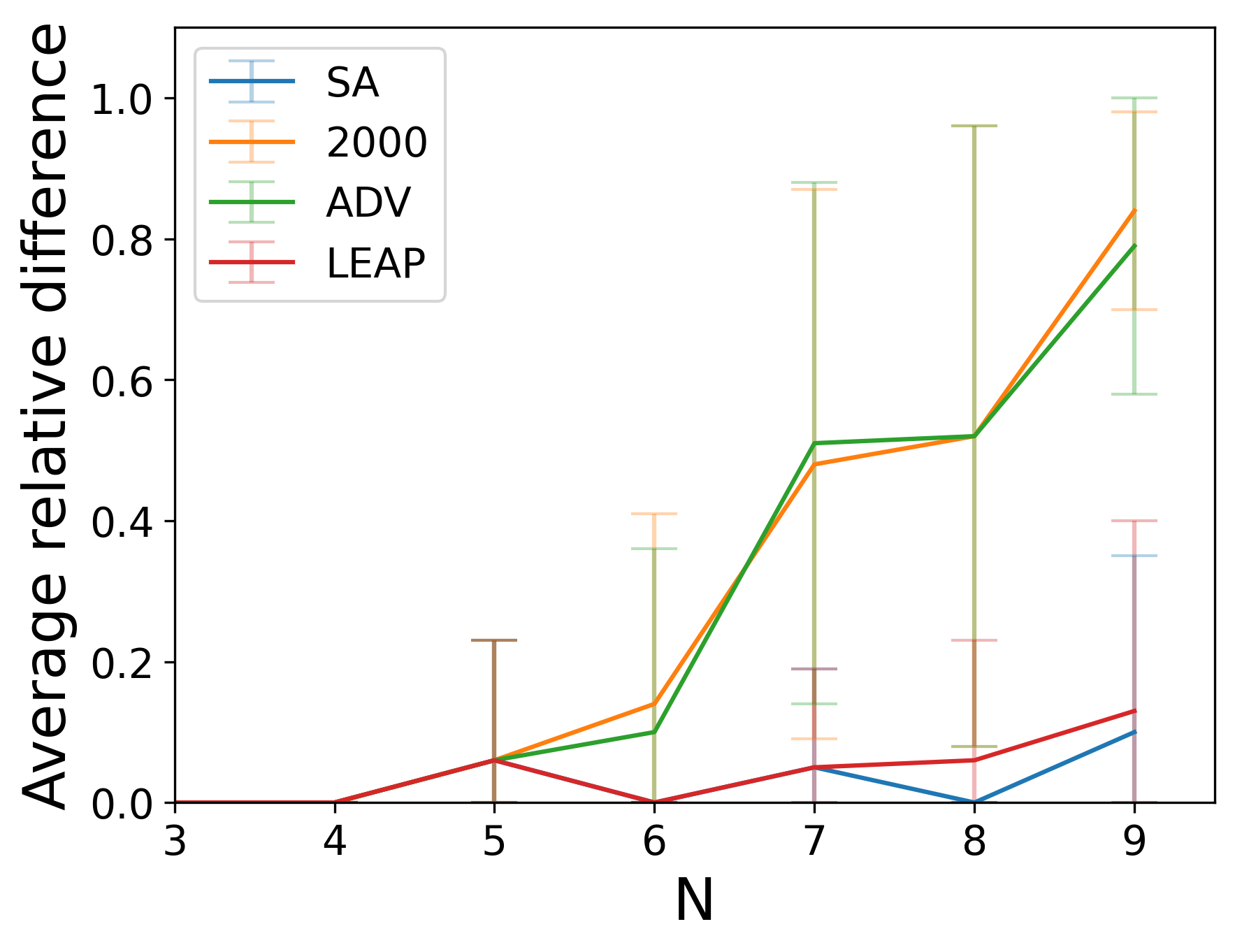}
        \caption{}
        \label{fig:result_dwave_best}
    \end{subfigure}~\begin{subfigure}[h]{0.26\textwidth}
        \centering
        \includegraphics[width=\textwidth]{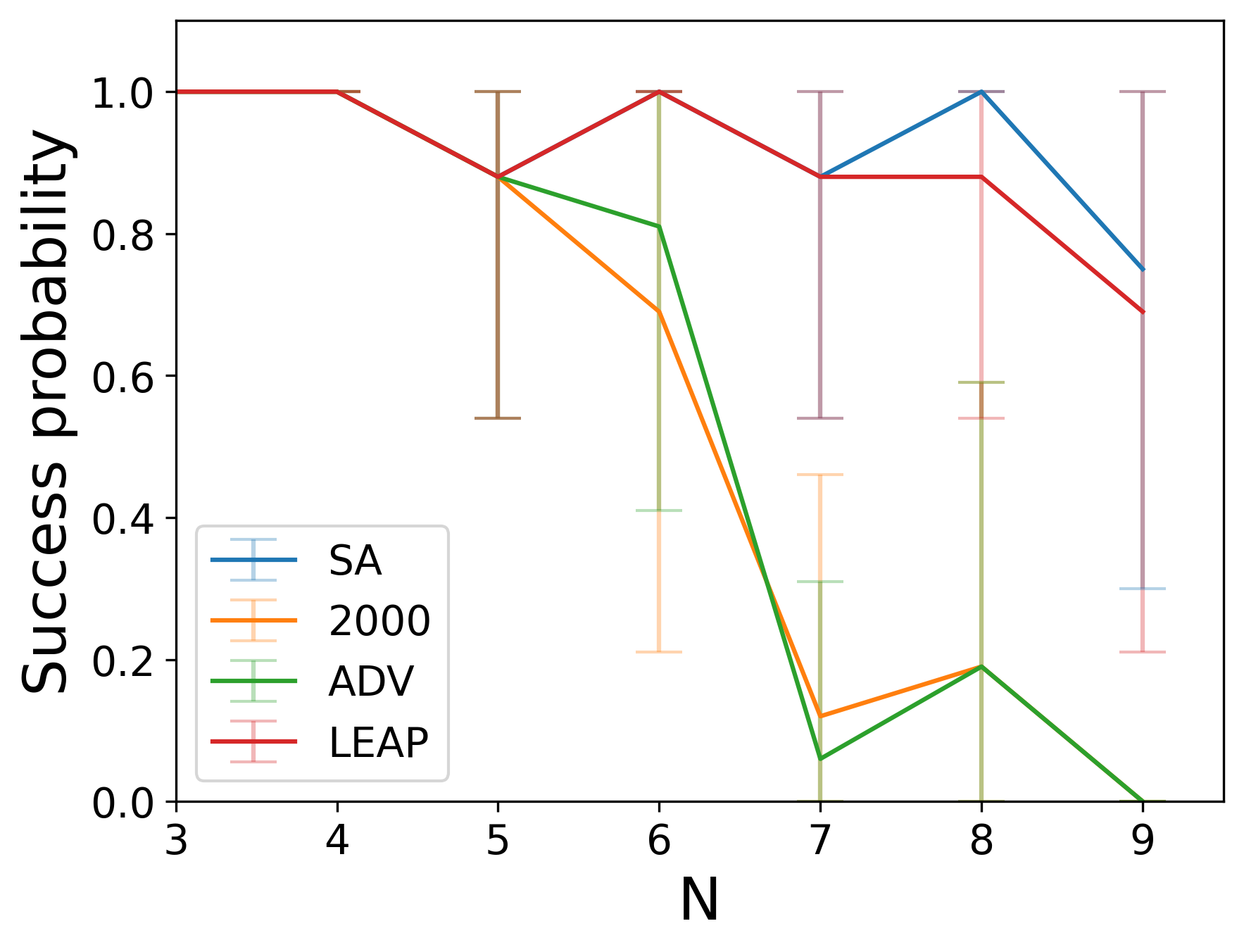}
        \caption{}
        \label{fig:success_prob_result_dwave_best}
    \end{subfigure}~\begin{subfigure}[h]{0.26\textwidth}
        \centering
        \includegraphics[width=\textwidth]{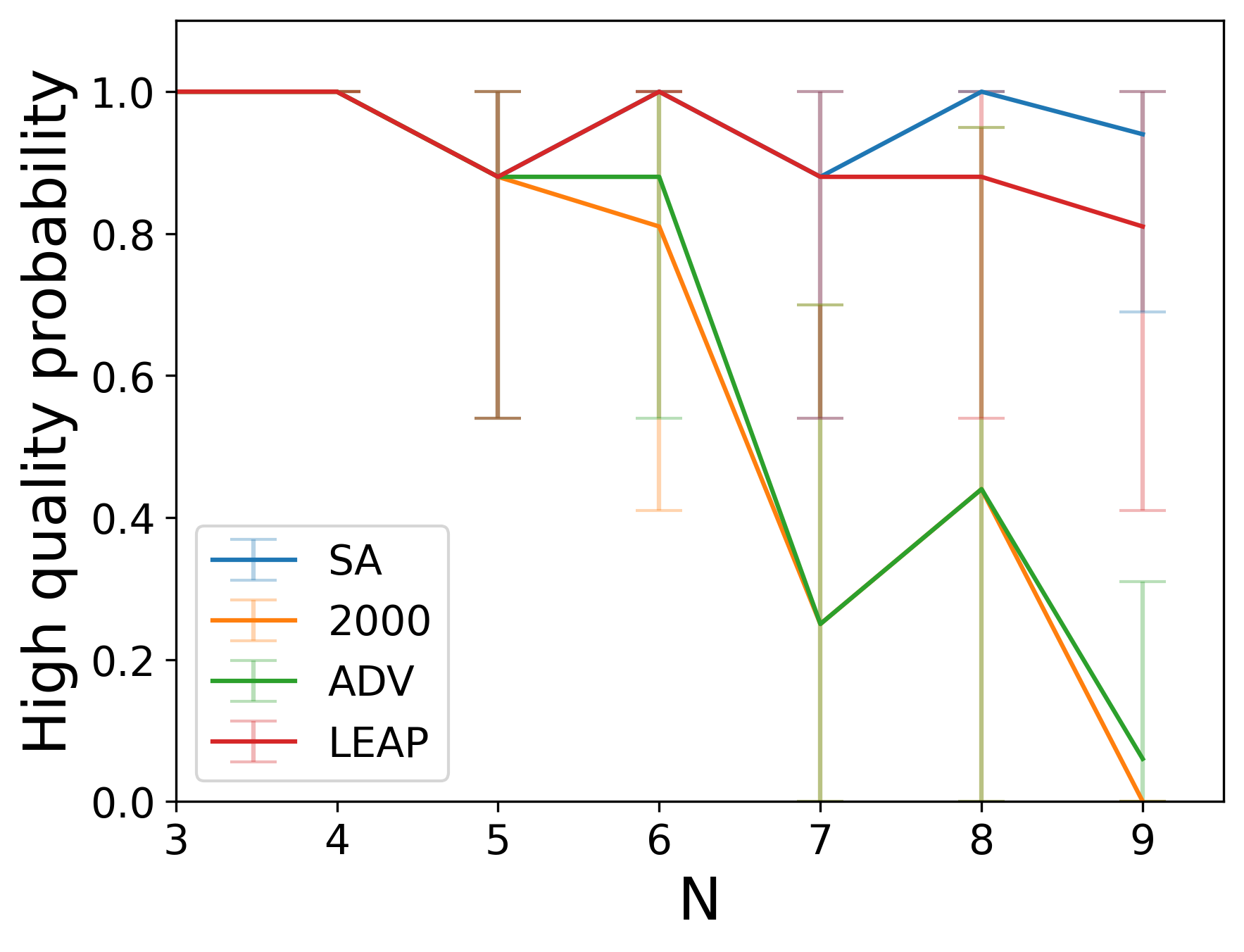}
        \caption{}
        \label{fig:hq_prob_result_dwave_best}
    \end{subfigure}
    \caption{
Comparison of the performance of SA and the different versions of quantum annealers. (a) shows the average relative difference, (b) shows the success probability, (c) shows the high quality probability. The error bars represent the standard deviation.
Legend:
SA is simulated annealing,
D2000 is D-Wave 2000Q having annealing time $20\si{\micro\second} \times 10^4$ runs,
ADV is D-Wave Advantage 1.1 having annealing time $20\si{\micro\second} \times 10^4$ runs,
LEAP is D-Wave Leap}
    \label{fig:best_annealer}
\end{figure*}

The resources exploited by the quantum annealer can give us useful information about the performance of our optimization tasks implemented over different configurations. We focus on the number of \emph{logical qubits,} i.e. the number of variables of the problem, and \emph{physical qubits}, i.e. the number of qubits needed to encode the problem by means of the minor embedding procedure.
We consider the \emph{maximum length of the chain}, that is the maximum number of physical qubits needed to encode a single logical qubit, and \emph{the chain strength} that measures the strength of the interaction between physical qubits belonging to the same chain.

As shown in Table~\ref{tab:resource_annealing}, the number of logical qubits depends uniquely on the graphs size. The other resources depend on both the topology of the quantum annealer and the quality of the software performing the minor embedding (which is the same for both versions of the hardware). It is evident that D-Wave Advantage produces much smaller embedding; as shorter chains lead to fewer errors, and in fact we obtain more accurate solutions. 

\begin{table}[tbp]
    \centering
    \begin{tabular}{l|rrrr|rrrr}
\toprule
  & \multicolumn{4}{c|}{D-Wave 2000} & \multicolumn{4}{c}{D-Wave Advantage} \\
$n$ &  L &    P & CL &   CS &  L &   P & CL &   CS \\\midrule
3 &  9 &   22 &  3 & 1.44 &  9 &  11 &  2 & 1.44 \\
4 & 16 &   65 &  5 & 1.76 & 16 &  33 &  3 & 1.76 \\
5 & 25 &  175 &  9 & 2.03 & 25 &  77 &  4 & 2.03 \\
6 & 36 &  377 & 14 & 2.27 & 36 & 159 &  6 & 2.27 \\
7 & 49 &  711 & 21 & 2.48 & 49 & 296 &  9 & 2.48 \\
8 & 64 & 1226 & 28 & 2.68 & 64 & 506 & 12 & 2.68 \\
9 & 81 & 1474 & 28 & 2.84 & 81 & 832 & 16 & 2.87 \\
\bottomrule
\end{tabular}
    \caption{Average resource usage for quantum annealers. Legend: 
    $n$: number of vertices, 
    L: logical qubits, 
    P: physical qubits, 
    CL: maximum chain length, 
    CS: chain strength}
    \label{tab:resource_annealing}
\end{table}

Then, we have quantified the resources required for variational algorithms. In Table~\ref{tab:resource_variational} we report the \emph{number of qubits} which is exactly to the number of variables. Since there is no need of minor embedding, the number of physical qubits corresponds to the number of logical ones. 
We also report the \emph{number of parameters} to be trained by the classical optimizer, the \emph{depth}, i.e. the maximum number of gates on one qubit,  and the \emph{size}, i.e. the total number of gates (see \cite{childs2019circuit} for a more formal definition of the notions of size and depth of a circuit).


To have a good estimation of resources, we transpile the circuit in terms of single-qubits rotations $U_3$ and CNOT gates.

\begin{table}[tbp]
    \centering
\scalebox{0.97}{\begin{tabular}{cr|rrr|rrr|rrr|rrr}
    \toprule
	& 
	& \multicolumn{6}{c|}{VQE}
	& \multicolumn{6}{c}{QAOA} \\
	& 
    & \multicolumn{3}{c}{$p=1$}
    & \multicolumn{3}{c|}{$p=3$}
    & \multicolumn{3}{c}{$p=1$}
    & \multicolumn{3}{c}{$p=3$}
    \\
    $n$ & Q 
    & P & D & S & P & D & S
    & P & D & S & P & D & S
    \\\midrule
3 & 9  & 36	 & 17	& 54	& 72  & 37	 & 144	 & 2 &  40 &   96 & 6 & 88  &   258 \\
4 & 16 & 64	 & 31	& 152	& 128 & 65	 & 424	 & 2 &  79 &  272 & 6 & 119 &   792 \\
5 & 25 & 100 & 49	& 350	& 200 & 101 & 1000	 & 2 & 125 &  638 & 6 & 187 &  1876 \\
6 & 36 & 144 & 71	& 702	& 288 & 145 & 2034	 & 2 & 183 & 1220 & 6 & 275 &  3606 \\
7 & 49 & 196 & 97	& 1274	& 392 & 197 & 3724	 & 2 & 257 & 2109 & 6 & 385 &  6255 \\
8 & 64 & 256 & 127	& 2144	& 512 & 257 & 6304	 & 2 & 329 & 3625 & 6 & 494 & 11259 \\
9 & 81 & 324 & 161	& 3402	& 648 & 325 & 10044  & 2 & 423 & 5535 & 6 & 635 & 16263 \\ \bottomrule
\end{tabular}}
    \caption{Average resource usage for variational algorithms. Legend: 
    $n$: number of vertices, 
    Q: number of qubits, 
    P: parameters, 
    D: depth, 
    S: size.}
    \label{tab:resource_variational}
\end{table}

We immediately see that in general VQE requires much more parameters than QAOA, although it depends mostly on the choice of the variational form. In general, this should lead to longer classical optimization phase. 

We were able to run experiments with graphs up to 5 vertices, since larger instances requires much more computational power for the simulation, which was not available to us.
The number of required gates suggests that this approach is not feasible on NISQ hardware due to the low gate fidelity and decoherence errors. Thus, we have performed the calculation on error-free simulators. 

\subsection{Comparing Quantum Annealers}

We identify the best performing configuration for all the quantum annealers, and then compare their performance with the simulated annealer.
Figure~\ref{fig:best_configuration} compares the different configuration of D-Wave 2000Q and Advantage 1.1. For both quantum annealer versions we see that:
\begin{itemize}
    \item configurations having 10000 runs perform significantly better than configurations with 1000 runs;
    \item the annealing time that minimizes the relative difference is $\tau \ge 20\si{\micro\second}$. Shorter values return inaccurate solutions, and larger values have similar performance but are costly. 
    \item the introduction of the pause in the annealing process does not improve the performance.
\end{itemize}

In Figure~\ref{fig:best_annealer} we compare the two versions of the quantum annealer, i.e. the D-Wave 2000Q and the D-Wave Advantage. We compared both versions in their best performing configuration, that is $20\si{\micro\second}$ of annealing time and $10^4$ runs, even though for some values of $n$ other configurations might slightly outperform this one. For most instances, the performance of D-Wave Advantage is better than the one of D-Wave 2000Q, but the gap between the two is quite small. 
The anomalous case ($n=7$) in which the average values of the metrics used seem to suggest that D-Wave 2000Q shows a better performance is related to numerical fluctuations. However, taking the experimental error into consideration, the conclusions drawn for the other cases are still valid. 
The few cases ($n=7$) where D-Wave 2000Q shows better performances are explained by the experimental error, as shown by the error bars. 
However, this fact slightly contrasts our expectation of better performance of D-Wave Advantage suggested by its more compact minor embedding (less physical qubits used).

Both quantum annealers perform worse than SA. Moreover, SA in general performs better than any other classical algorithm tested (results in Appendix~\ref{appendix:results}). We can state the superior performances of classical hardware compared to the current quantum annealers, for the GED problem.

It is important to notice that Hybrid Annealer D-Wave Leap has performance close to the SA, even without ever outperforming it. The significance of this finding is weakened by the opaque internal working of D-Wave Leap, which might be perform SA itself.

Finally, we have identified which configuration promises the best tradeoff between quality of the solution and total annealing time. This observation is relevant because the cost of using a quantum annealer is proportional to the amount of time used by the machine.

We measure the \emph{time to solution} (TTS) as 
\[ \mathrm{TTS} = \frac{\text{\# of runs} \times \text{annealing time per run}}{\text{high quality probability}} \]
for both D-Wave 2000 and D-Wave Advantage. It measures how much time in seconds is required on a quantum annealer to find a high quality solution. In the literature you can find also different definitions of TTS (\cite{ttsdef1, ttsdef2}), all representing the same concept.
The results are shown in Table~\ref{tab:tts}. Our experiments show that having the shortest annealing time with a large number of runs gives the best tradeoff in terms of TTS. For $n=9$ vertices no experiment found any high quality solution thus was not possible to estimate such probability. 

\begin{table}[htbp]
    \centering
    \begin{tabular}{c|rrrrr|rrrrr}
        \toprule
          & \multicolumn{5}{c|}{D-Wave 2000Q} & \multicolumn{5}{c}{D-Wave Advantage} \\
        $n$ & A & B & C & D & E & A & B & C & D & E \\\midrule
        3 & 0.01 & 0.20 & 0.02 & 0.50 & 0.10 & 0.01 & 0.20 & 0.02 & 0.50 & 0.10 \\
        4 & 0.01 & 0.20 & 0.02 & 0.50 & 0.10 & 0.01 & 0.20 & 0.02 & 0.50 & 0.10 \\
        5 & 0.01 & 0.23 & 0.02 & 0.57 & 0.12 & 0.01 & 0.23 & 0.02 & 0.57 & 0.11 \\
        6 & 0.03 & 0.25 & 0.04 & 0.72 & 0.14 & 0.01 & 0.23 & 0.04 & 0.57 & 0.18 \\
        7 & 0.05 & 0.80 & 0.08 & 1.61 & 0.32 & 0.03 & 0.80 & 0.11 & 2.00 & 0.32 \\
        8 & 0.17 & 0.45 & 0.33 & 1.14 & 0.32 & 0.02 & 0.45 & 0.17 & 8.33 & 1.67 \\
        \bottomrule
    \end{tabular}
    \caption{Time To Solution for the two hardware configurations, measured in seconds. Legend: 
    $n$: number of vertices, 
    A: $1\si{\micro\second}$ per run $\times 10k$ runs,
    B: $20\si{\micro\second}$ per run $\times 10k$ runs,
    C: $20\si{\micro\second}$ per run $\times 1k$ runs,
    D: $500\si{\micro\second}$ per run $\times 1k$ runs,
    E: $100\si{\micro\second}$ per run $\times 1k$ runs with paused annealing.
    }
    \label{tab:tts}
\end{table}

\subsection{Comparing variational algorithms}

\begin{figure}[htbp]
    \centering
    \begin{subfigure}[h]{0.13\textwidth}
        \centering
        \includegraphics[width=\textwidth]{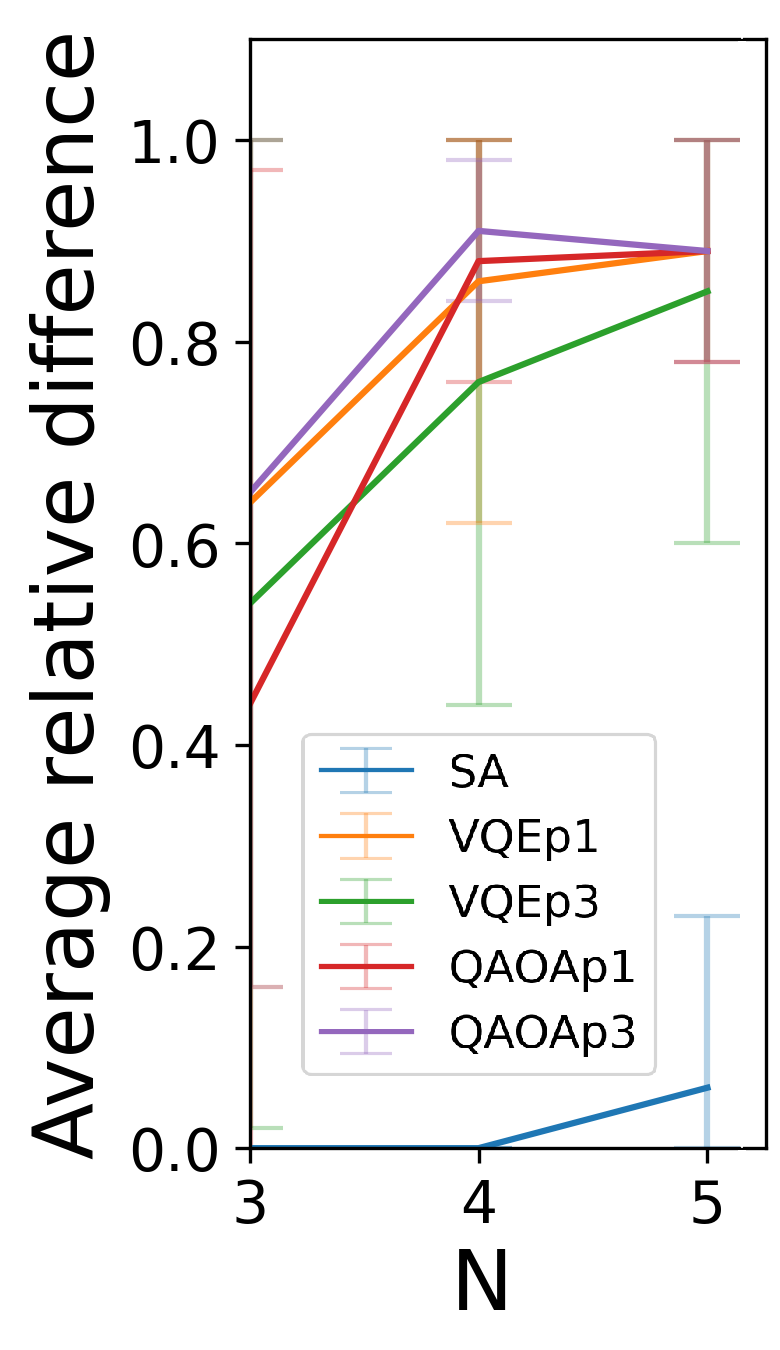}
        \caption{}
        \label{fig:result_variational}
    \end{subfigure}\quad%
    \begin{subfigure}[h]{0.13\textwidth}
        \centering
        \includegraphics[width=\textwidth]{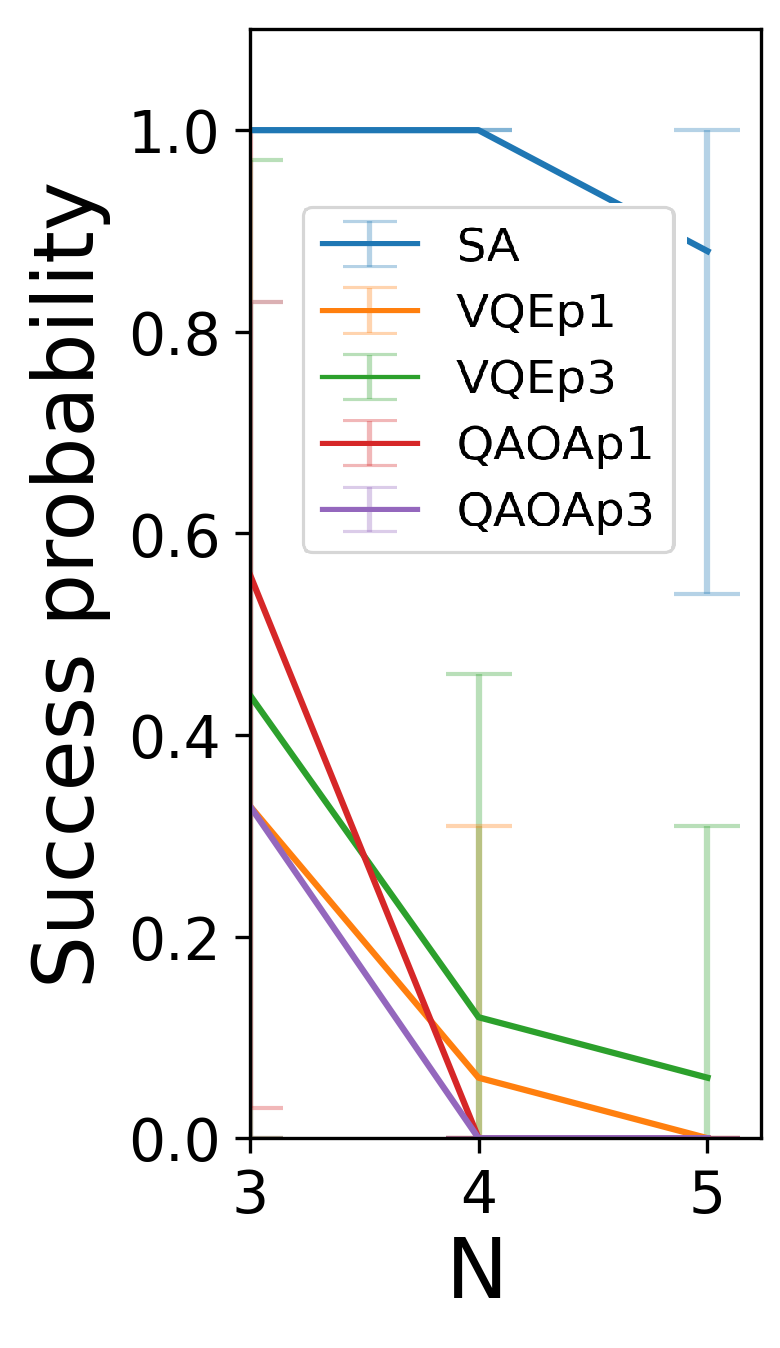}
        \caption{}
        \label{fig:success_prob_result_variational}
    \end{subfigure}\quad%
    \begin{subfigure}[h]{0.13\textwidth}
        \centering
        \includegraphics[width=\textwidth]{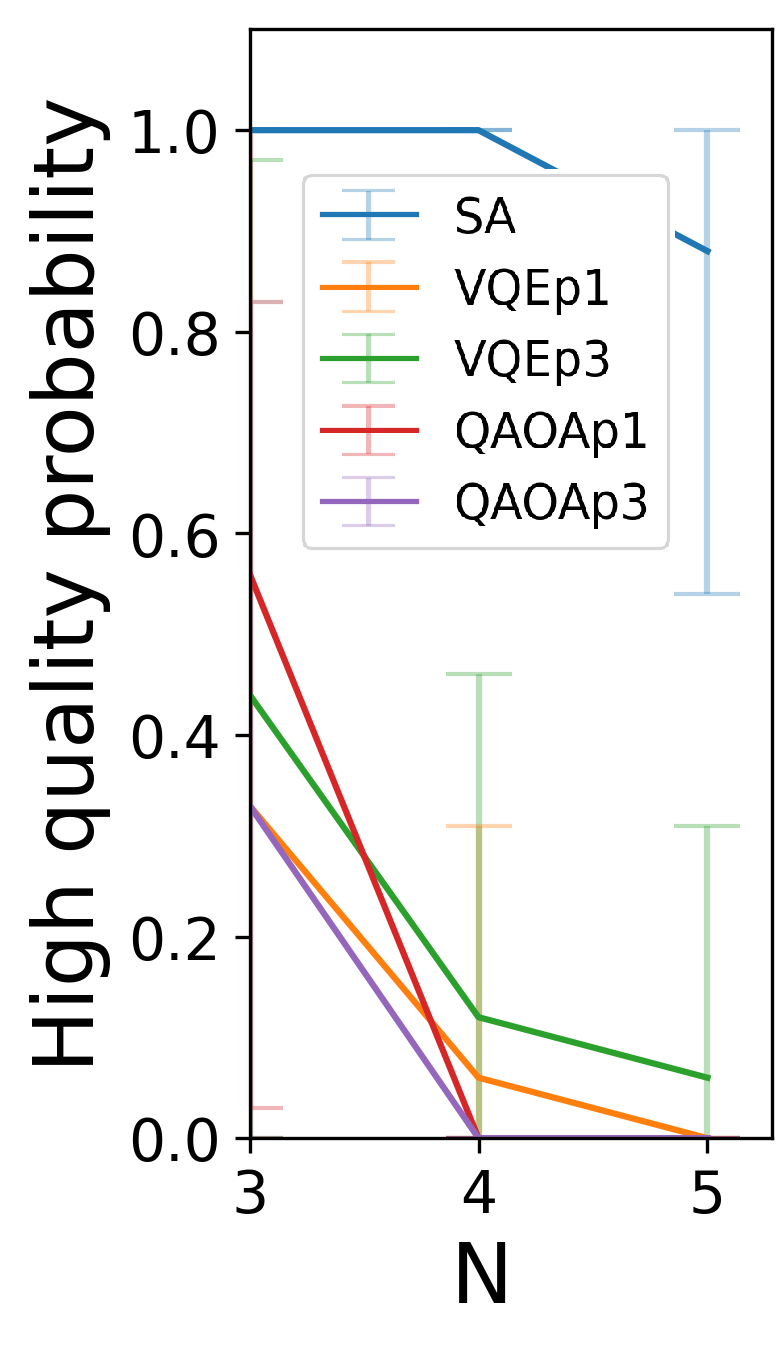}
        \caption{}
        \label{fig:hq_prob_result_variational}
    \end{subfigure}
    \caption{Performance of variational algorithms to compute the GED for graphs with $n$ vertices. In (\ref{fig:result_variational}) we plot the mean relative difference, while in (\ref{fig:success_prob_result_variational}) we plot the success solution probability and in (\ref{fig:hq_prob_result_variational}) the high quality solution probability. The acronyms refer to the different algorithms as follows (the error bars represent the standard deviation): V1 is VQE  with $p=1$, V3 is VQE with $p=3$,     Q1 is QAOA with $p=1$, and Q3 is QAOA  with $p=3$.}
    \label{fig:results_variational}
\end{figure}

Figure~\ref{fig:results_variational} shows the comparison of the performance of VQE, QAOA, and SA. Considering the current state of NISQ devices, we have tested the variational approach on instances of GED with graphs having at most 5 vertices (thus our circuit uses 25 qubits). For these small dimensions, the classical approach leads to much better solutions. In particular, no variational approaches are able to find an exact solution for instances having $n \ge 4$ vertices. 

Increasing the number of repetitions may marginally improve the performances: VQE with $p=3$ outperforms VQE with $p=1$ for any choice of $n$, while for QAOA the configuration with $p=1$ performs equal or better than the one with $p=3$.
Theoretically, in the limit of $p \rightarrow \infty$, the probability of success is  guaranteed to be 1 (see \cite{fahriqaoa}). However, increasing the number of layers becomes really costly in terms of 
the evaluation of the expectation value of the cost Hamiltonian. In fact, the dimensions of the parameter search space scales, when no {\it a priori} symmetries are known, 
as $[- \pi, \pi]^p \times [- \pi, \pi]^p $. To gain insight on the low performances of the QAOA algorithm we consider the energy landscape of the associate Hamiltonian of the QUBO formulation of the GED of two graphs having four vertices and with $p=1$. 
For the case considered, the Hamiltonian $H_C$ has a discrete spectrum with 88 eigenenergies, whose ground state is degenerate.

\begin{figure}
    \includegraphics[width=0.90\columnwidth]{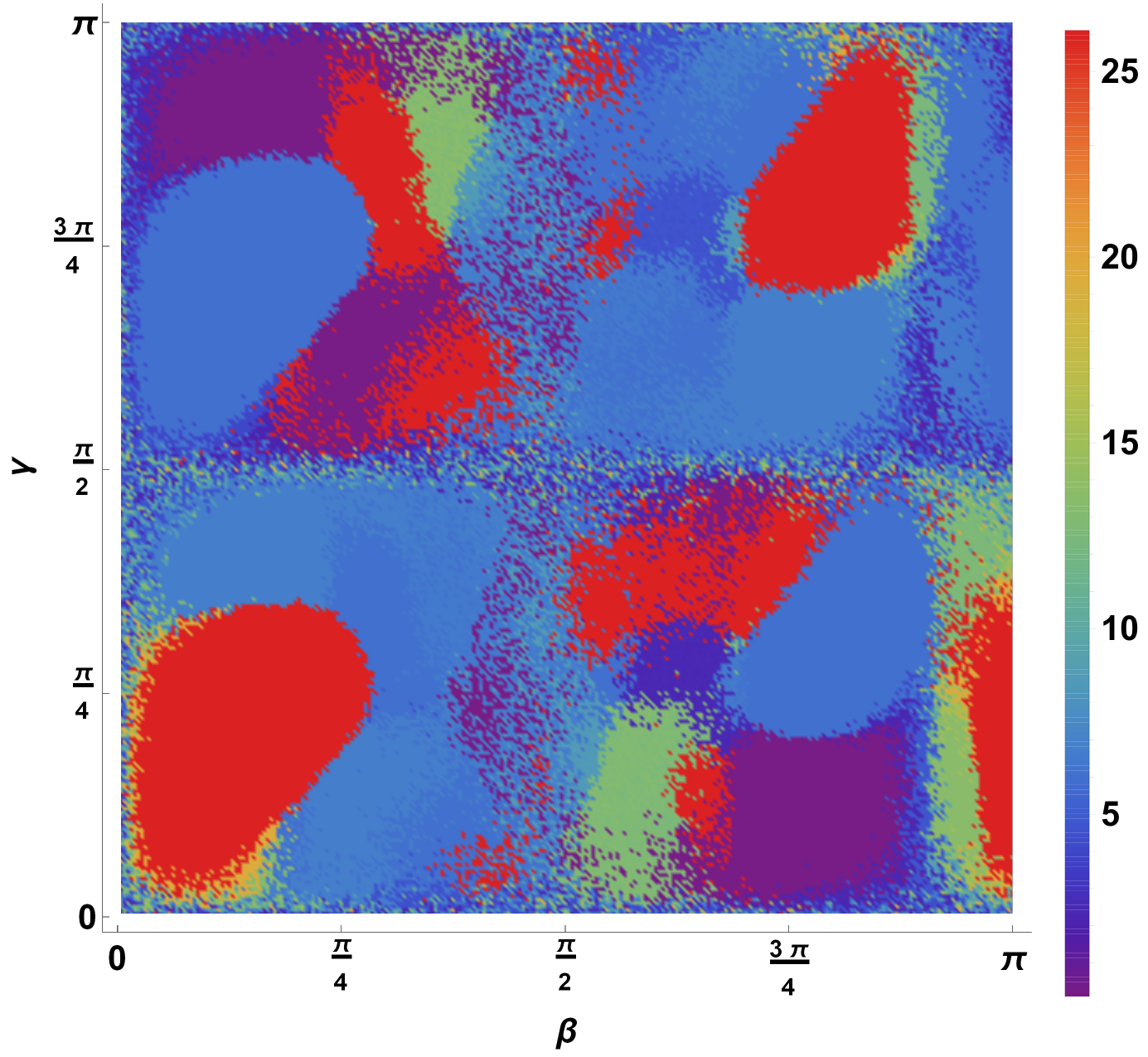}
	\caption{\label{fig:energy_Qaoa}
   Energy Landscape given by \ref{energy_qaoa} for a Hamiltonian referring to a pair of graphs with $N=4$. Due to the symmetry of the system we have plotted the landscape only in the relevant range $[0, \pi] \times [0, \pi]$.}
\end{figure}

We report the energy landscape obtained in Fig. \ref{fig:energy_Qaoa}, where it is clear that the Hamiltonian does not have a unique global minima, leading to a low performance of the optimization technique provided by the QAOA for the QUBO formulation of the GED problem.

\section{Concluding remarks}\label{sec:conclusions}

Complex systems are nowadays ubiquitous in science.
Very useful models for these systems are typically defined by representing them as graphs, i.e. collections of pairwise connected nodes. The nodes constitute the elementary `units' of the problem, 
while the edges take into account their interaction. 
Each edge can also be associated with a label representing some 
quantitative or qualitative information.
This versatility of the graph structures makes them a powerful tool in the most heterogeneous fields of research. 
 
In this paper we have addressed the problem of quantitatively estimating  the degree of similarity between a pair of graphs via the \emph{Graph Edit Distance}. 
Computing GED is a task that requires the exploration of a space of solutions that is exponential in the size of the input graphs. It is therefore reasonable to consider approximate approaches to the problem, which are able to achieve acceptable approximations of the exact solution. We have thus investigated  whether the resources offered by the quantum hardware currently available are plausible candidates to tackle this task. 
Our proof-of-principle analysis has had a twofold aim: on one side we have shown how to practically implement the computation of GED on both a quantum annealer and a gate-based quantum computer, and subsequently we have compared the results of running the same GED algorithm on the the two types of hardware. Based on our results, the quantum annealer seems to be today a better platform for optimization problems  written as QUBO problems. 

We remark that, concerning the variational approach, we have used noiseless simulators to have a first insight on the feasibility of evaluating the GED with NISQ devices. Nowadays, the development of the gate-based quantum hardware as well as the specific software for optimization algorithms is still in its infancy. In this work we have reported some preliminary evidences that the main variational algorithms may not be suited for the type of problems as the one we have addressed in this paper. Overall, we are confronted with two contrasting aspects of the currently available resources. On one hand, classically simulating a quantum computer is expensive although it does not suffer from problems related to quantum physical resources, namely qubits and unitary gates acting for a sufficiently long time \cite{Unit_long}. On the other hand these quantum computational resources are very limited so as to make it impossible to exploit their full potential. However, we surmise that these detrimental effects will be overcome when quantum hardware with enough resources to allow error-correction schemes as a barrier against noise will be available. Another problem with variational algorithms is that it is necessary to guess the right encoding of the problem into the parameterized cost function that is evaluated using a quantum computer. This is a challenging but crucial task for the success of the solution scheme, together with the choice of the best classical optimizer for the parameter training phase.


In this direction, it was recently proposed a Quantum Natural Gradient Descent algorithm \cite{stokes2020quantum, gacon2021simultaneous}, which seems to offer enhanced performance by considering statistical information about the quantum circuit such as geometrical methods based on the Quantum Fisher Information \cite{abbas2021power}. 
Another seemingly promising strategy is to combine the two approaches that we have studied in this paper, namely using the quantum annealing to get preliminary results that can then be used to initialize a variational algorithm, as shown for the case of the QAOA in \cite{Sack2021quantumannealing}. 

We believe that the algorithms and the benchmarking of quantum annealers and gate-based quantum computers that we have presented in this paper can also be exploited for machine learning tasks. In fact, the algorithms we have devised in our implementations could be profitably incorporated in machine learning algorithms that deal with graph data, to obtain quantum algorithms that are more efficient than the standard classical machine learning strategies. 



\section*{Acknowledgements}

We would like to thank CINECA for making available to us the necessary resources for running our tests (QPU D-Wave and classical supercomputers for simulations), through the Italian SuperComputing Resource Allocation - ISCRA - programme. 
A.M. acknowledges the University of Verona for financial support throughout Mobility Grant No. INTAT20DIPIERRO, the Foundation for Polish Science (FNP)
for the IRAP project ICTQT (Contract No. 2018/MAB/5),
cofinanced by the EU Smart Growth Operational Programme, and the Polish National Science Center for the grant MINIATURA DEC-2020/04/X/ST2/01794.




%


\onecolumngrid
\appendix

\clearpage
\section{Algorithms}
\label{appendix:algo}
In this Appendix we present the algorithms that we have implemented for creating our benchmark. 

Algorithm~\ref{alg:reduction} is used to construct the QUBO matrix that represents the QUBO formulation of the GED problem. 

\begin{algorithm}[H]
\caption{Construct QUBO matrix}\label{alg:reduction}
\begin{algorithmic}
\Require $G, H$ graphs having the same number $|n|$ of vertices
\Ensure $Q$ $n \times n$ real matrix
    \State $Q \gets$ $n \times n$ symmetric matrix initialized to zero \Comment{$Q[a,b]$ denotes both the $a$-th row, $b$-th col and $a$-th col, $b$-th row since the matrix is symmetric}
    \State $\lambda_h \gets n^2 + 1$
    \State $\lambda_s \gets 1$
    \State \Comment{construct hard constraint (squared formula expanded}
    \ForAll{$i \in [0, n)$} \Comment{row hard constraint}
        \ForAll{$j \in [0, n)$} \Comment{$x_{i,j}^2 - 2x_{i,j}$}
            \State $\ell \gets i\times n + j$
            \State $Q[\ell, \ell] \gets Q[\ell, \ell] - \lambda_h$ 
        \EndFor
        \ForAll{$j \in [0, n)$} \Comment{$2 x_{i,j} x_{i,j'}$}
            \ForAll{$j' \in [j, n)$}
                \State $\ell \gets i\times n + j$
                \State $\ell' \gets i\times n + j'$
                \State $Q[\ell, \ell'] \gets Q[\ell, \ell'] + 2 \lambda_h$ 
            \EndFor
        \EndFor
    \EndFor
    \ForAll{$j \in [0, n)$} \Comment{column hard constraint}
        \ForAll{$i \in [0, n)$} \Comment{$x_{i,j}^2 - 2x_{i,j}$}
            \State $\ell \gets i\times n + j$
            \State $Q[\ell, \ell] \gets Q[\ell, \ell] - \lambda_h$ 
        \EndFor
        \ForAll{$i \in [0, n)$} \Comment{$2 x_{i,j} x_{i',j}$}
            \ForAll{$i' \in [j, n)$}
                \State $\ell \gets i\times n + j$
                \State $\ell' \gets i'\times n + j$
                \State $Q[\ell, \ell'] \gets Q[\ell, \ell'] + 2 \lambda_h$ 
            \EndFor
        \EndFor
    \EndFor
    \ForAll{$i,j \in E(G)$} \Comment{$P$ constraints}
        \ForAll{$i' \in [0, n)$}
            \ForAll{$j' \in [0, n)$}
                \If{$i',j' \not\in E(H)$}
                    \State $\ell \gets i\times n + i'$
                    \State $\ell' \gets j\times n + j'$
                    \State $Q[\ell, \ell'] \gets Q[\ell, \ell'] + \lambda_s$ 
                \EndIf
            \EndFor
        \EndFor
    \EndFor
    \ForAll{$i',j' \in E(H)$} \Comment{$Q$ constraints}
        \ForAll{$i \in [0, n)$}
            \ForAll{$j \in [0, n)$}
                \If{$i,j \not\in E(G)$}
                    \State $\ell \gets i\times n + i'$
                    \State $\ell' \gets j\times n + j'$
                    \State $Q[\ell, \ell'] \gets Q[\ell, \ell'] + \lambda_s$ 
                \EndIf
            \EndFor
        \EndFor
    \EndFor
    \State \textbf{return} $Q$
\end{algorithmic}
\end{algorithm}

Algorithm~\ref{alg:SA} represents how the Simulating Annealing procedure we have used works. 

\begin{algorithm}[H]
\caption{Simulated annealing}\label{alg:SA}
\begin{algorithmic}[1]
\Require $Q$ QUBO matrix sized $n \times n$
\Require $T_0 > 0$ initial temperature
\Require $\alpha \in (0,1)$ temperature decrease factor
\Require $S > 0$ number of samples
\Ensure $E$ optimal or sub-optimal cost for minimizing QUBO

\State $\mathrm{samples} \gets$ empty list
\State $\mathrm{energies} \gets$ empty list
\Repeat
    \State Generate $x \in \{0,1\}^n$ randomly
    \State $T \gets T_0$ \Comment{initial temperature}
    \State $E \gets x^T Q x$ \Comment{initial energy}
    \Repeat
        \State $x' \gets \text{slightly perturbed }x$ 
        \State $E' \gets (x')^T Q (x')$
        \State $\Delta E \gets E' - E$
        \State $r \gets$ random value between 0 and 1 
        \If{$\Delta E \le 0$} \Comment{accept the solution}
            \State $x \gets x'$
            \State $E \gets E'$
        \ElsIf{$e^{-(\Delta E)/T} > r$} \Comment{accept the solution}
            \State $x \gets x'$
            \State $E \gets E'$
        \Else
            \State do nothing  \Comment{reject solution}
        \EndIf
        \State $T \gets \alpha \cdot T$ \Comment{decrease temperature}
    \Until{$T \approx 0$} 
    \State append $x$ to $\mathrm{samples}$
    \State append $E$ to $\mathrm{energies}$
    \State $S \gets S - 1$
\Until{$S = 0$} \Comment{On D-Wave's implementation you can define also a non-zero final temperature}

\State \textbf{return} $\mathrm{samples}, \mathrm{energies}$ \Comment{alternatevely, return $\min(\mathrm{energies})$}
\end{algorithmic}
\end{algorithm}

Algorithm~\ref{alg:vqeansatz} constructs the parametric quantum circuit needed to run the VQE algorithm. 

\begin{algorithm}[H]
\caption{Construct VQE circuit}\label{alg:vqeansatz}
\begin{algorithmic}[1]
\Require $p$ repetition of the ansatz
\Require $\theta \in [-\pi, \pi]^{2pn}$ parameters vector
\Ensure Circuit for VQE
\State Create circuit of $n$ qubits
\For{$i \in 1, ..., p$}
    \For{$j \in 1, ..., n$}
        \State Apply gate $R_y(\theta_{2in+2i})$ on $j$-th qubit
        \State Apply gate $R_y(\theta_{2in+2i+1})$ on $j$-th qubit
    \EndFor
    \For{$j \in 1, ..., n-1$}
        \For{$k \in i+1, ..., n$}
            \State Apply gate $CX$ on $j$-th (control) and $k$-th (target) qubits
        \EndFor
    \EndFor
\EndFor
\State \textbf{return} circuit
\end{algorithmic}
\end{algorithm}

Algorithm~\ref{alg:qaoaansatz} constructs the parametric quantum circuit needed to run the QAOA algorithm.

\begin{algorithm}[H]
\caption{Construct QAOA circuit}\label{alg:qaoaansatz}
\begin{algorithmic}[1]
\Require $H_C = \sum_{i,j} J_{i,j} \sigma^z_i \sigma^z_j$ defined on $n$ variables
\Require $p$ number of repetition of the Hamiltonians
\Require $\gamma \in [-\pi, \pi]^p$ parameter vector
\Require $\beta \in [-\pi, \pi]^p$ parameter vector
\Ensure Circuit for QAOA

\State Create circuit of $n$ qubits \Comment{Circuit starts in state $\ket{0}^{\otimes n}$}
\For{$i \in 1, ..., n$} \Comment{Circuit evolves in state $\ket{+}^{\otimes n}$}
    \State Apply gate $H$ on $i$-th qubit
\EndFor
\For{$i \in 1, ..., p$}
    \For{$J_{i,j} \sigma^z_i \sigma^z_j$ in $H_c$} \Comment{Construct $U_C(\gamma_i) = e^{-i \gamma_i H_C}$}
        \If{$i=j$}
            \State Apply gate $R_{Z}(\gamma_i J_{i,j})$ on $i,j$-th qubits
        \Else
            \State Apply gate $R_{ZZ}(\gamma_i J_{i,j})$ on $i,j$-th qubits
        \EndIf
    \EndFor
    \For{$i \in 1, ..., n$} \Comment{Construct $U_B(\gamma_i) = e^{-i \beta_i H_B}$}
        \State Apply gate $R_{X}(\beta_i)$ on $i$-th qubit
    \EndFor
    \State \textbf{return} circuit
\EndFor

\end{algorithmic}
\end{algorithm}

\clearpage
\section{Detailed results}\label{appendix:results}

\begin{figure}[htbp]
    \centering
    \includegraphics[width=0.85\textwidth]{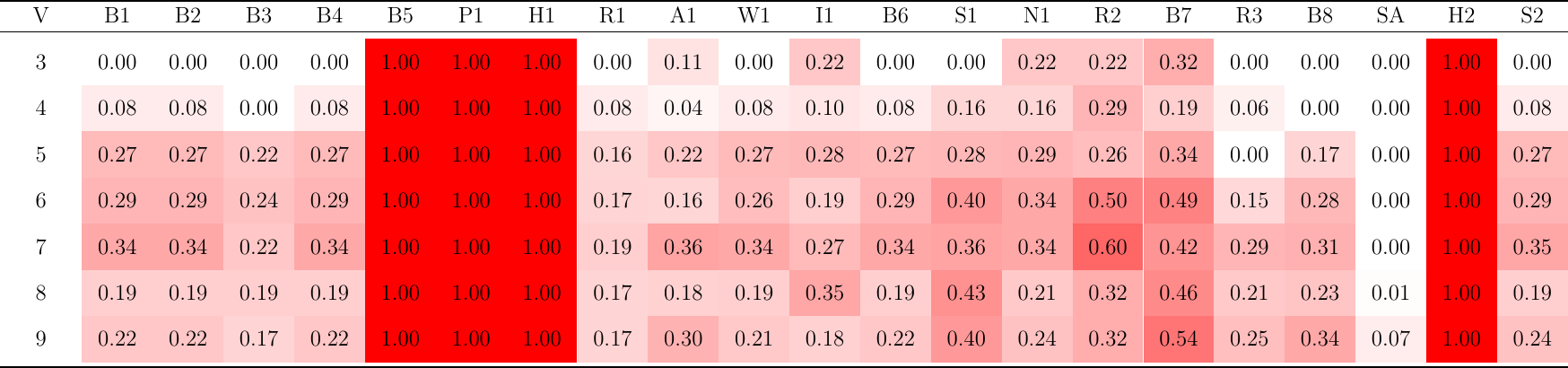}
    \caption{Average relative difference (lower is better), Legend: 
V: number of vertices,
B1: Branch,
B2: Branch Fast,
B3: Branch Tight,
B4: Branch Uniform,
B5: Branch Compact,
P1: Partition,
H1: Hybrid,
R1: Ring,
A1: ANCHOR\_AWARE\_GED,
W1: Walks,
I1: IPFP,
B6: BIPARTITE,
S1: SUBGRAPH,
N1: NODE,
R2: RING\_ML,
B7: BIPARTITE\_ML,
R3: REFINE,
B8: BP\_BEAM,
SA: Simulated Annealing (GEDLib implementation, not D-Wave's)
H2: HED,
S2: STAR}
    \label{fig:classical_mean_difference_probability}
\end{figure}

\begin{figure}[htbp]
    \centering
    \includegraphics[width=0.85\textwidth]{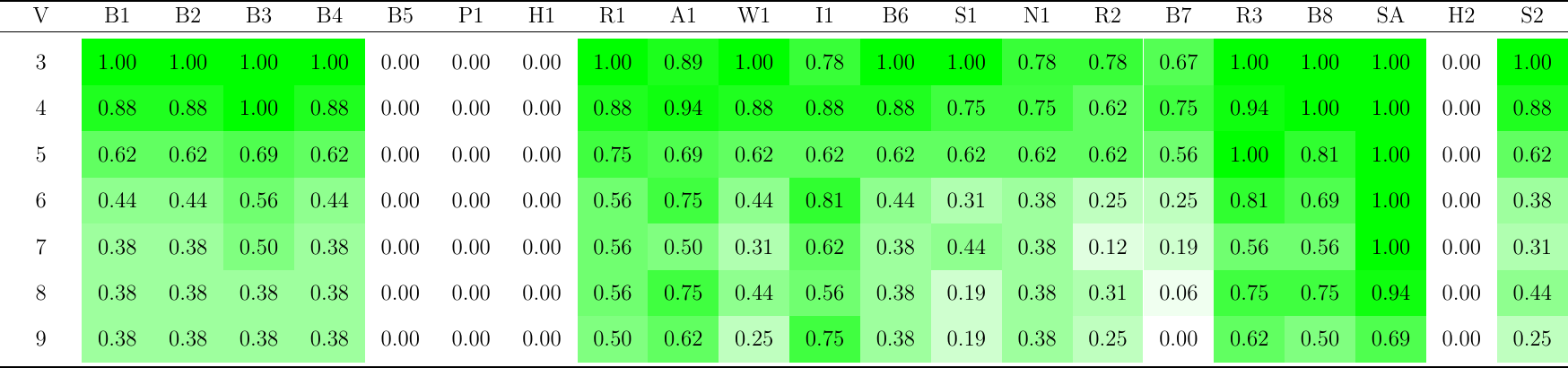}
    \caption{Average Success probability (higher is better), Legend: 
V: number of vertices,
B1: Branch,
B2: Branch Fast,
B3: Branch Tight,
B4: Branch Uniform,
B5: Branch Compact,
P1: Partition,
H1: Hybrid,
R1: Ring,
A1: ANCHOR\_AWARE\_GED,
W1: Walks,
I1: IPFP,
B6: BIPARTITE,
S1: SUBGRAPH,
N1: NODE,
R2: RING\_ML,
B7: BIPARTITE\_ML,
R3: REFINE,
B8: BP\_BEAM,
SA: Simulated Annealing (GEDLib implementation, not D-Wave's)
H2: HED,
S2: STAR}
    \label{fig:classical_success_probability}
\end{figure}

\begin{figure}[htbp]
    \centering
    \includegraphics[width=0.85\textwidth]{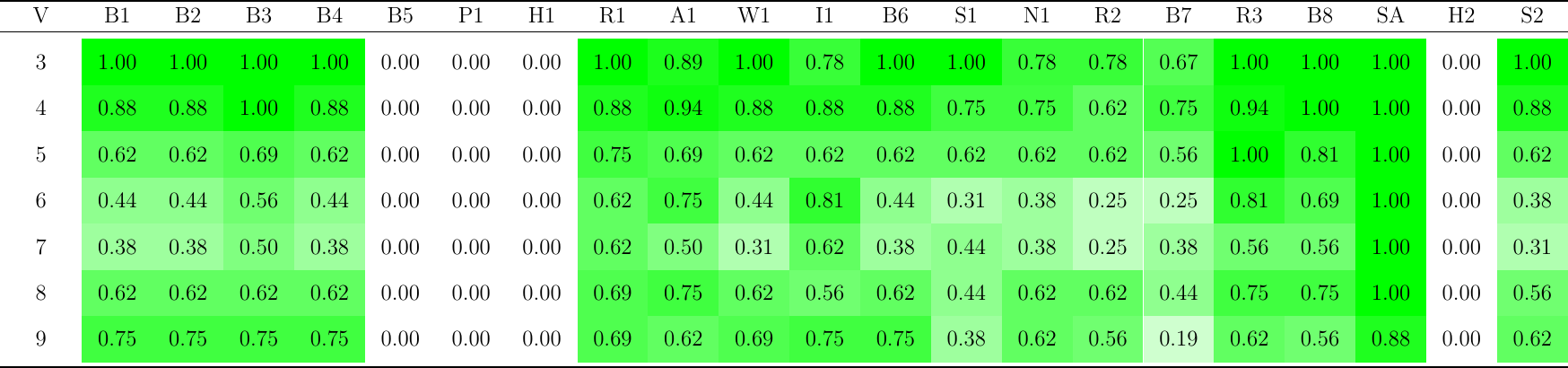}
    \caption{Average High quality ($<20\%$) probability (higher is better), Legend: 
V: number of vertices,
B1: Branch,
B2: Branch Fast,
B3: Branch Tight,
B4: Branch Uniform,
B5: Branch Compact,
P1: Partition,
H1: Hybrid,
R1: Ring,
A1: ANCHOR\_AWARE\_GED,
W1: Walks,
I1: IPFP,
B6: BIPARTITE,
S1: SUBGRAPH,
N1: NODE,
R2: RING\_ML,
B7: BIPARTITE\_ML,
R3: REFINE,
B8: BP\_BEAM,
SA: Simulated Annealing (GEDLib implementation, not D-Wave's)
H2: HED,
S2: STAR}
    \label{fig:classical_hq_probability}
\end{figure}

Figure~\ref{fig:classical_mean_difference_probability}-\ref{fig:classical_hq_probability} compares the performances of classical heuristics. All these approaches are detailed explained in \cite{BlumenthalGED}.

Figure~\ref{fig:classical_mean_difference_probability} contains the Average relative difference, as defined in Section~\ref{sec:methods} by Equation~\ref{eq:relativedifference}.

Figure~\ref{fig:classical_success_probability} contains the Average success probability, as defined in Section~\ref{sec:numerical_results}, which is the percentage of runs having null relative difference over all runs.

Figure~\ref{fig:classical_hq_probability} contains the Average high-quality probability, as defined in Section~\ref{sec:numerical_results}, which is the percentage of experiments having results with relative difference $\leq 0.2$ .

\begin{figure}[htbp]
    \centering
    \includegraphics[width=0.70\textwidth]{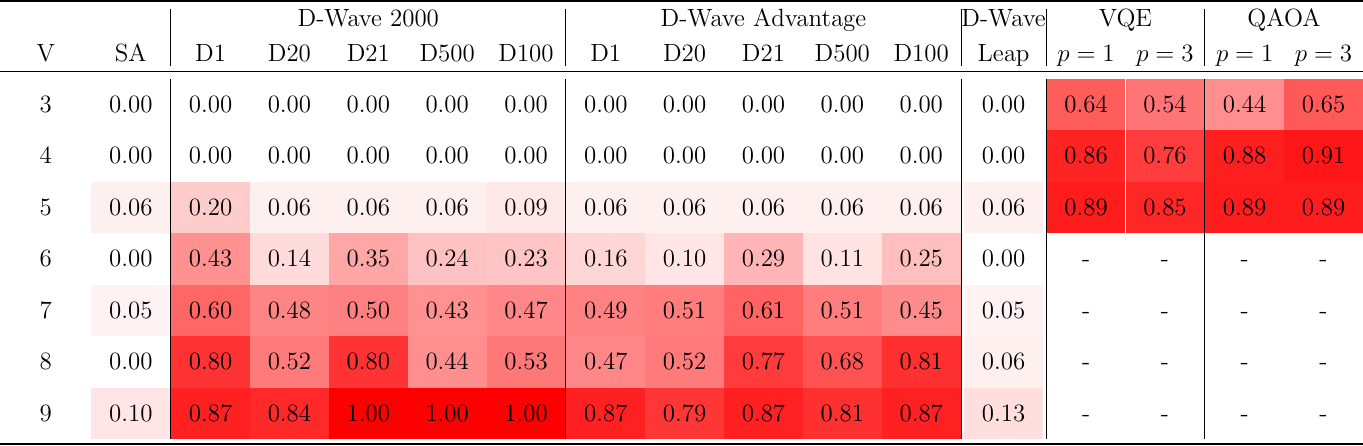}
    \caption{Average relative difference (lower is better). Legend: 
    V: number of vertices, 
    SA: Simulated Annealing (D-Wave's implementation),
    D1: annealing $1\si{\micro\second} \times 10^4$ runs, 
    D20: annealing $20\si{\micro\second} \times 10^4$ runs, 
    D21: annealing $20\si{\micro\second} \times 10^3$ runs, 
    D500: annealing $500\si{\micro\second} \times 10^3$ runs, 
    D100: annealing $100\si{\micro\second}$ paused in the middle, $\times 10^3$ runs,
    $p$: number of repetition of the variational form}
    \label{fig:quantum_mean_difference_probability}
\end{figure}

\begin{figure}[htbp]
    \centering
    \includegraphics[width=0.70\textwidth]{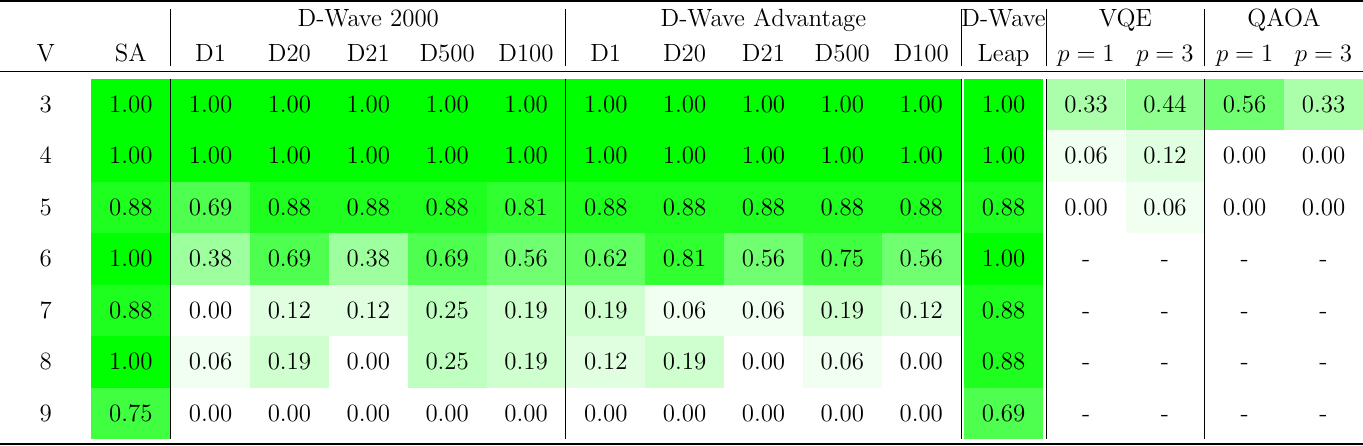}
    \caption{Average success probability (higher is better).  Legend: 
    V: number of vertices, 
    SA: Simulated Annealing (D-Wave's implementation),
    D1: annealing $1\si{\micro\second} \times 10^4$ runs, 
    D20: annealing $20\si{\micro\second} \times 10^4$ runs, 
    D21: annealing $20\si{\micro\second} \times 10^3$ runs, 
    D500: annealing $500\si{\micro\second} \times 10^3$ runs, 
    D100: annealing $100\si{\micro\second}$ paused in the middle, $\times 10^3$ runs,
    $p$: number of repetition of the variational form}
    \label{fig:quantum_success_probability}
\end{figure}

\begin{figure}[htbp]
    \centering
    \includegraphics[width=0.70\textwidth]{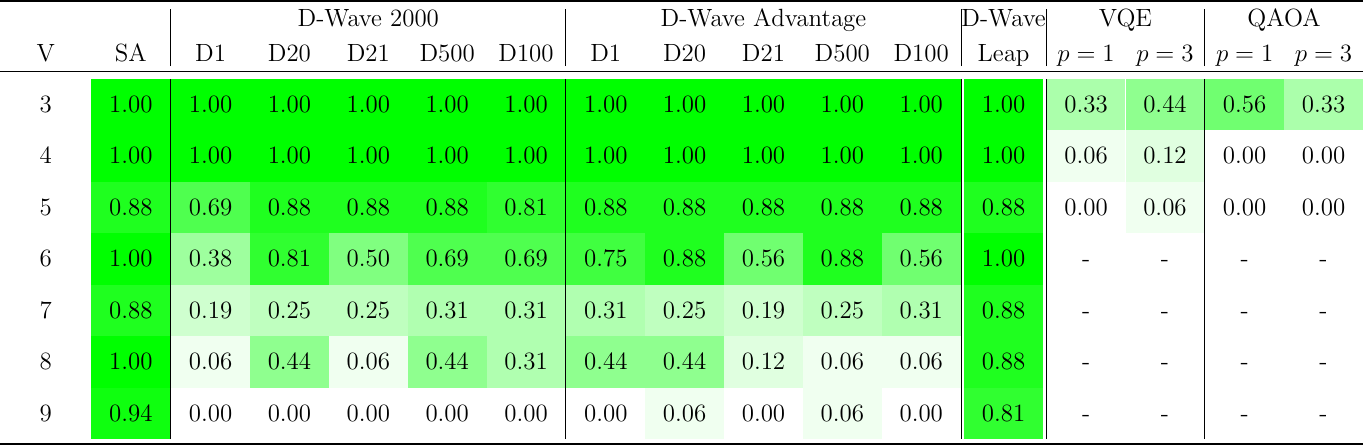}
    \caption{Average high quality probability (higher is better).  Legend: 
    V: number of vertices, 
    SA: Simulated Annealing (D-Wave's implementation),
    D1: annealing $1\si{\micro\second} \times 10^4$ runs, 
    D20: annealing $20\si{\micro\second} \times 10^4$ runs, 
    D21: annealing $20\si{\micro\second} \times 10^3$ runs, 
    D500: annealing $500\si{\micro\second} \times 10^3$ runs, 
    D100: annealing $100\si{\micro\second}$ paused in the middle, $\times 10^3$ runs,
    $p$: number of repetition of the variational form}
    \label{fig:quantum_hq_probability}
\end{figure}

\clearpage Figure~\ref{fig:quantum_mean_difference_probability}-\ref{fig:quantum_hq_probability} compares the performances of simulated annealing, quantum annealing, variational algorithms.

Figure~\ref{fig:quantum_mean_difference_probability} contains the Average relative difference, as defined in Section~\ref{sec:methods} by Equation~\ref{eq:relativedifference}.

Figure~\ref{fig:quantum_success_probability} contains the Average success probability, as defined in Section~\ref{sec:numerical_results}, which is the percentage of runs having null relative difference over all runs.

Figure~\ref{fig:quantum_hq_probability} contains the Average high-quality probability, as defined in Section~\ref{sec:numerical_results}, which is the percentage of experiments having results with relative difference $\leq 0.2$.

\end{document}